\documentclass[prb,twocolumn,floatfix,superscriptaddress]{revtex4-1}
\usepackage[pdftex,plainpages=false,colorlinks=true,linkcolor=blue, citecolor=blue, urlcolor=blue]{hyperref}
\usepackage{amsfonts}
\usepackage{amsmath}
\usepackage{amssymb}
\usepackage{graphicx}
\usepackage{natbib}
\usepackage{color}
\usepackage{placeins}

\begin{document}
 
\title{Effects of interaction strength, doping, and frustration on the antiferromagnetic phase of the two-dimensional Hubbard model}
\author{L. Fratino}
\affiliation{Department of Physics, Royal Holloway, University of London, Egham, Surrey, UK, TW20 0EX}
\author{M. Charlebois}
\affiliation{D\'epartement de physique, Institut quantique, and Regroupement Qu\'eb\'ecois sur les mat\'eriaux de Pointe, Universit\'e de Sherbrooke, Sherbrooke, Qu\'ebec, Canada J1K 2R1}
\author{P. S\'emon}
\affiliation{Computational Science Initiative, Brookhaven National Laboratory, Upton, NY 11973-5000, USA}
\author{G. Sordi}
\affiliation{Department of Physics, Royal Holloway, University of London, Egham, Surrey, UK, TW20 0EX}
\author{A.-M. S. Tremblay}
\affiliation{D\'epartement de physique, Institut quantique, and Regroupement Qu\'eb\'ecois sur les mat\'eriaux de Pointe, Universit\'e de Sherbrooke, Sherbrooke, Qu\'ebec, Canada J1K 2R1}
\affiliation{Canadian Institute for Advanced Research, Toronto, Ontario, Canada, M5G 1Z8}

\date{\today}

\begin{abstract}
Recent quantum-gas microscopy of ultracold atoms and scanning tunneling microscopy of the cuprates reveal new detailed information about doped Mott  antiferromagnets, which can be compared with calculations. Using cellular dynamical mean-field theory, we map out the antiferromagnetic (AF) phase of the two-dimensional Hubbard model as a function of interaction strength $U$, hole doping $\delta$ and temperature $T$. The N\'eel phase boundary is non-monotonic as a function of $U$ and $\delta$. Frustration induced by second-neighbor hopping reduces N\'eel order more effectively at small $U$. The doped AF is stabilized at large $U$ by kinetic energy and at small $U$ by potential energy. The transition between the AF insulator and the doped metallic AF is continuous. At large $U$, we find in-gap states similar to those observed in scanning tunneling microscopy. We predict that, contrary to the Hubbard bands, these states are only slightly spin polarized. 
\end{abstract}
 
\maketitle

The quantum mechanics of interacting electrons on a lattice can lead to complex many-body phase diagrams. For example, doping a layered Mott insulator can give rise to antiferromagnetism, pseudogap, unconventional superconductivity and multiple exotic phases~\cite{keimerRev}. 
The Hubbard model is the simplest model of interacting electrons on a lattice. It can be used for both natural (e.g. cuprates) and artificial (e.g. ultracold atoms) systems~\cite{Anderson:1987, jzHM, AMJulich, AntoineLesHouches}. 
Therefore understanding the phases that appear in this model and the transitions between them is a central programme in condensed matter physics. 

Here we study the regimes where antiferromagnetic (AF) correlations set in within the two dimensional (2D) Hubbard model on a square lattice as a function of interaction $U$, doping $\delta$ and temperature $T$, within cellular dynamical mean-field theory (CDMFT)~\cite{maier, kotliarRMP, tremblayR}. The motivation for our work is threefold. 
First, recent advances in ultracold atom experiments can now reach temperatures low enough to detect AF correlations for repulsively interacting Fermi gases~\cite{Greif:2013, Hart:2015, Parsons:2016, Boll:2016, Cheuk:2016, mazurenko2017cold, drewesPRL2017}. Hence, a theoretical characterisation of the AF phase in the whole $U-\delta-T$ space might guide ultracold atom experiments that are exploring this uncharted territory. 
Second, recent tunneling spectroscopy studies~\cite{CaiSTM, YeSTM} reveal new details on the evolution of the AF Mott insulator upon doping, thus calling for theoretical explanations. 
Third, on the theory side we still know little about the detailed boundaries of the AF phase in the whole $U-\delta-T$ space of the 2D Hubbard model and the mechanism by which AF is stabilized. Most previous studies with this and other methods focused on $T=0$~\cite{Dupuis2004, senechalAFSC2005, markus2005, markus, Tocchio2016, ZhengDMET}. The negative sign problem hampers the study of finite $T$, large $U$ and finite doping~\cite{Hirsch:1985, White:1989, lkAF, Paiva:2001, maier, kyung, Paiva2010, Sato2016}. Our results might serve as a stepping stone for new approaches directed towards including Mott physics and long wavelength fluctuations~\cite{trilex1, trilex2, quadrilex, Rohringer2017}.

\begin{figure*}[ht!]
\centering{
\includegraphics[width=0.99\linewidth]{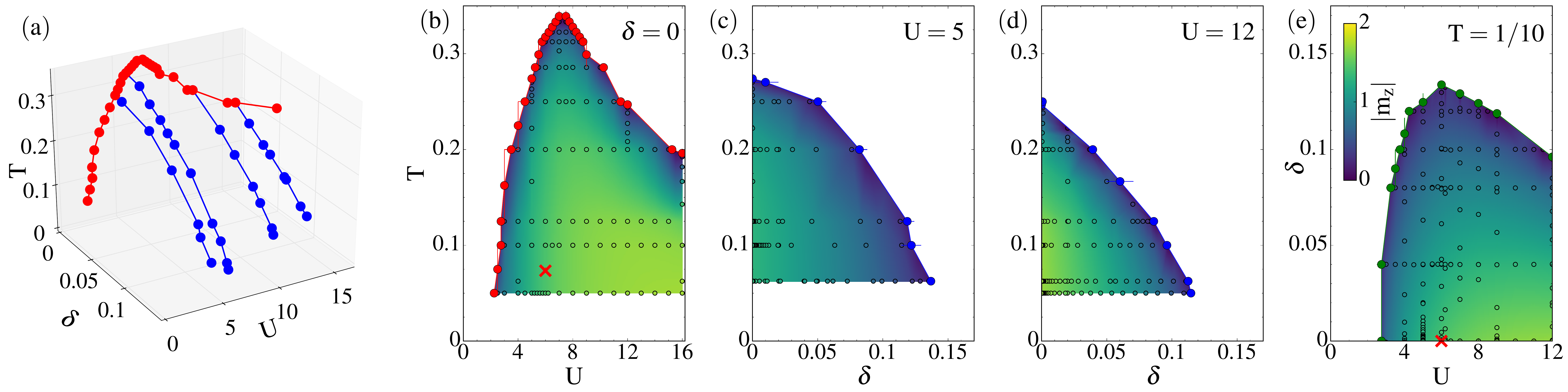}
}
\caption{(a) AF phase of the 2D Hubbard model in the $U-T-\delta$ space, for $t'=-0.1$. (b) $T-U$ cut at $\delta=0$. (c) $T-U$ cut for $U=5$ and (d) $U=12$. (e) $\delta-U$ cut for $T=1/10$. The magnitude of $m_z$ is color coded [see SM~\cite{Note1} for $m_z(U)$ and $m_z(\delta)$ curves at different temperatures]. The paramagnetic to AF phase boundary is drawn where $m_z <0.045$. Red cross in panels (b) and (c) indicates the position of the Mott critical endpoint $(U_{\rm MIT}, T_{\rm MIT})$ in the underlying normal phase. 
}
\label{fig1}
\end{figure*}
%


{\it Model and method. --}  
We consider the 2D Hubbard model, $H=-\sum_{ij\sigma}t_{ij}c_{i\sigma}^\dagger c_{j\sigma}
  +U\sum_{i} n_{i\uparrow } n_{i\downarrow }
  -\mu\sum_{i\sigma} n_{i\sigma}$, 
where $t_{ij}=t~(t')$ is the (next) nearest-neighbor hopping, $U$ is the onsite Coulomb repulsion and $\mu$ is the chemical potential. Here $c^{\dagger}_{i\sigma}$ ($c_{i\sigma}$) is the creation (destruction) operator on lattice site $i$ and spin $\sigma$, and $n_{i\sigma}$ is the number operator. We set $t=1$ as our energy unit. Within the cellular extension~\cite{maier, kotliarRMP, tremblayR} of dynamical mean-field theory~\cite{rmp}, a $2\times 2$ plaquette is embedded in a self-consistent bath. We have successfully benchmarked this approach~\cite{LorenzoAF2017} at $\delta=0$, where reliable results are available. 

We solve the cluster impurity problem using continuous time Quantum Monte Carlo based on the expansion of the hybridization between impurity and bath~\cite{millisRMP, patrickSkipList}. 
Symmetry breaking is allowed only in the bath. It is efficient to use of the $C_{2v}$ group symmetry with mirrors along plaquette diagonals~\cite{LorenzoAF2017, patrickCritical, patrickERG}.

{\it $U-T-\delta$ map of the AF phase. --} 
Long-wavelength spin fluctuations lead, in two dimensions, to a vanishing staggered magnetization $m_z$ at finite temperature~\cite{MWtheorem,Hohenberg:1967}. Nevertheless, $m_z = \frac{2}{N_c} \sum_i(-1)^i (n_{i\uparrow} - n_{i\downarrow})$ is non-zero in cold-atom experiments because of finite-size effects. For cuprates, the $m_z$ that we compute becomes non-vanishing at a dynamical mean-field N\'eel temperature $T^d_N$ where the antiferromagnetic correlation length of the infinite system would start to grow exponentially~\cite{LorenzoAF2017}. Coupling in the third dimension leads to true long-range order at a lower temperature.  

As a first step, $m_z$ is used to map out the AF phase in the $U-T-\delta$ space for $t'=-0.1$. We consider hole doping only ($\delta=1-n>0$) and perform various cuts, i.e. (i) at $\delta=0$ ($T-U$ plane in Fig.~\ref{fig1}b), (ii) at fixed values of $U$ ($T-\delta$ planes in Figs.~\ref{fig1}c,d), (iii) and at fixed temperature $T$ ($\delta-U$ plane at $T=1/10$ in Fig.~\ref{fig1}e). These cuts are reported in the $U-T-\delta$ space in Fig.~\ref{fig1}a, where one sees that $T_N^d(U,\delta)$ has a global maximum at $U \approx 7$ for $\delta=0$. The sign problem prevents convergence below $T\approx 1/20$. The value of $m_z \neq 0$ is color coded in Figs.~\ref{fig1}b-e and shown in the Supplemental Material (SM)~\footnote{See Supplemental Material for the staggered magnetization curves as a function of $U$ and $\delta$ for few temperatures; complementary data for the local DOS.}.

The staggered magnetization $m_z(U,\delta, T)$ is largest in the $\delta=0$ plane and saturates for large $U$ and low $T$, as in mean field~\cite{Zhang:1989}.  
Our analysis of the AF region in the $U-T-\delta$ space highlights two points. First, the overall behavior of $m_z$ differs from that of $T_N^d$: For example, $m_z(U,T=0)|_{\delta=0}$ does not scale with either $T_N^d(U)|_{\delta=0}$ (phase boundary in Fig.~\ref{fig1}b), or with $\delta_N^d(U)|_T$ (phase boundary in Fig.~\ref{fig1}e). 
Physically even if large $U$ creates local moments, $T_N^d$ decreases with $U$ since it is the superexchange $J=4t^2/U$ that aligns these moments at finite temperature. Second, the maxima~\cite{GeorgesKrauthAFM:1993,FreericksJarrelAFM:1995} of both $T_N^d(U)|_{\delta=0}$ and $\delta_N^d(U)|_T$ are correlated with the Mott transition that exists at $\delta=0$ in the unstable normal state below $T_N^d$, suggesting that the hidden Mott transition [see Mott endpoint in Fig.\ref{fig1}b,e] drives the qualitative changes in the AF state. 

It is well known that the increase of $T_N^d(U)$ at small $U$ is explained by the Slater physics of nesting and that the decrease of $T_N^d(U)$ at large $U$ is explained by the Heisenberg physics of superexchange. Hence, the fact that the position of the maximum of $T_N^d(U)$ at $\delta=0$ is controlled by the underlying Mott transition~\cite{LorenzoAF2017} reflects the underlying physics. As we saw above (cf. green line in Fig.~\ref{fig1}e), this difference between small and large $U$ persists upon doping since the range of $\delta$ where AF exists first increases with $U$ and then decreases, with the crossover again controlled by the Mott transition at $\delta=0$.  In contrast, regardless of the strength of $U$, $T_N^d(\delta)$ monotonically decreases with increasing $\delta$ [phase boundaries in Figs.~\ref{fig1}c,d].

\begin{figure*}[ht!]
\centering{
\includegraphics[width=0.99\linewidth]{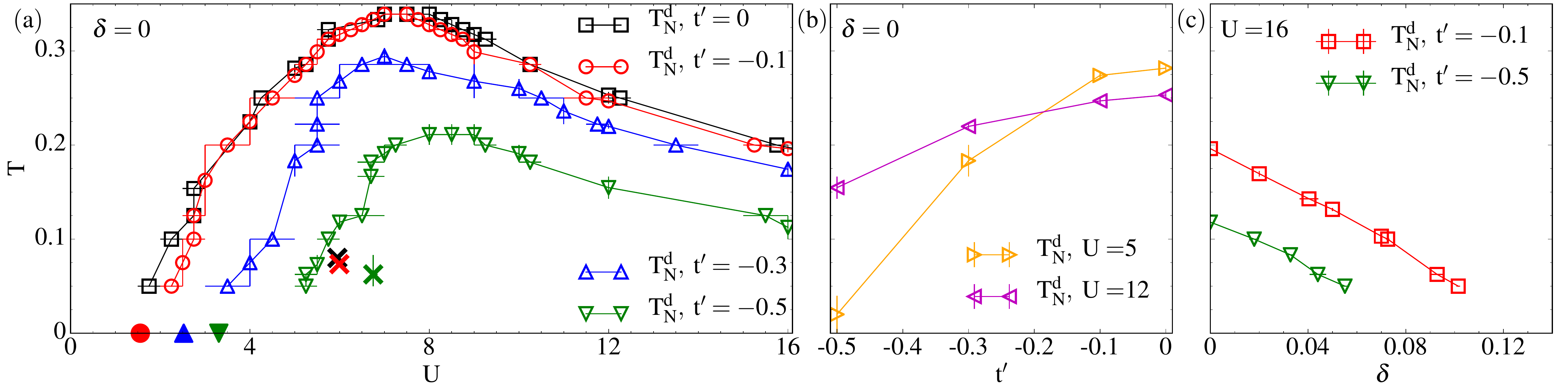}
}
\caption{(a) $T_N^d$  for $t'=0,-0.1,-0.3,-0.5$. As reference, we show with filled symbols the $T=0$ Hartree-Fock results of Ref.~\onlinecite{hofstetter1998} for the critical $U$ of the AF onset. Crosses indicate the Mott endpoints in the underlying normal phase: data are for $t'=0$ (black), $t'=-0.1$ (red) and $t'=-0.5$ (green). (b) $T_N^d$ versus $t'$ at $\delta=0$ for $U=5,12$. (c) $T_N^d$ versus $\delta$ at $U=16$ for $t'=-0.1,-0.5$. }
\label{fig2}
\end{figure*}

{\it Effect of frustration on $T_N^d(U,\delta)$. --}
We can gain further insights by varying the next-nearest neighbor hopping $t'$, which frustrates AF order in varying degree depending on the value of $U$, as we shall see. Having in mind the physics of hole doped cuprates, here we consider only negative values of $t'$, in the range $t' \in [0, -0.5]$. 

Figure~\ref{fig2}a shows $T_N^d(U)$ at $\delta=0$ for different values of $t'$. 
AF now appears at a critical $U_c$ that shifts to higher values of $U$ upon increasing $|t'|$, in agreement with expectation from the physics of nesting and also from the $T=0$, DMFT $d=\infty$~\cite{ChitraAFM:1999,ZitzlerPruschkeAFM:2004} and Hartree-Fock (HF) results~\cite{hofstetter1998}. The $T\rightarrow 0$ transition at $U_c$ is consistent with first order~\cite{hofstetter1998,ChitraAFM:1999,ZitzlerPruschkeAFM:2004} for finite $t'$ (see Fig.~2 in SM~\cite{Note1}). $U_c$ is larger than the HF result~\cite{hofstetter1998} because the vertex is renormalized downward compared to the bare $U$ by fluctuations in other channels~\cite{kanamori_electron_1963,Brueckner:1960,Vilk:1997, kotliarSBaf}. We find once again, that the position of max~$T_N^d(U)$ is correlated with the Mott transition in the underlying normal state. 

Although frustration reduces $T_N^d(U)$ as expected, the reduction of $T_N^d(U)$ upon increasing $|t'|$ at $\delta=0$ is stronger at small $U$ than at large $U$, as shown in Fig.~\ref{fig2}b. Indeed, although at small $U$ deviations from perfect nesting are first order in $|t'/t|$, at large $U$ the AF arises from localized spins and the correct quantities to compare are $J'=4 t'^2/U$ and $J=4 t^2/U$ whose ratio scale as $|t'/t|^2$. 
Figure~\ref{fig2}c shows the doping-dependent $T_N^d(\delta)$ at $U=16$ for different values of $t'$: at our lowest temperature, a  fivefold increase of $|t'|$ only approximately halves the critical doping $\delta_N^d$ at which the AF phase ends. The robustness of the AF phase at finite $\delta$ seems to reflect the robustness at $\delta=0$ since we observe a rigid downward shift of the whole $T_N^d(\delta)$ line. The transition at the critical $\delta$ is consistent with second order (SM~\cite{Note1} Fig.~1d).

{\it AF insulator to AF metal transition. --}
Having mapped out the N\'eel state, we next explore its nature by analyzing the local density of states (DOS) $N(\omega)$ and the occupation $n(\mu)=1-\delta(\mu)$. 

\begin{figure*}
\centering{
\includegraphics[width=0.99\linewidth]{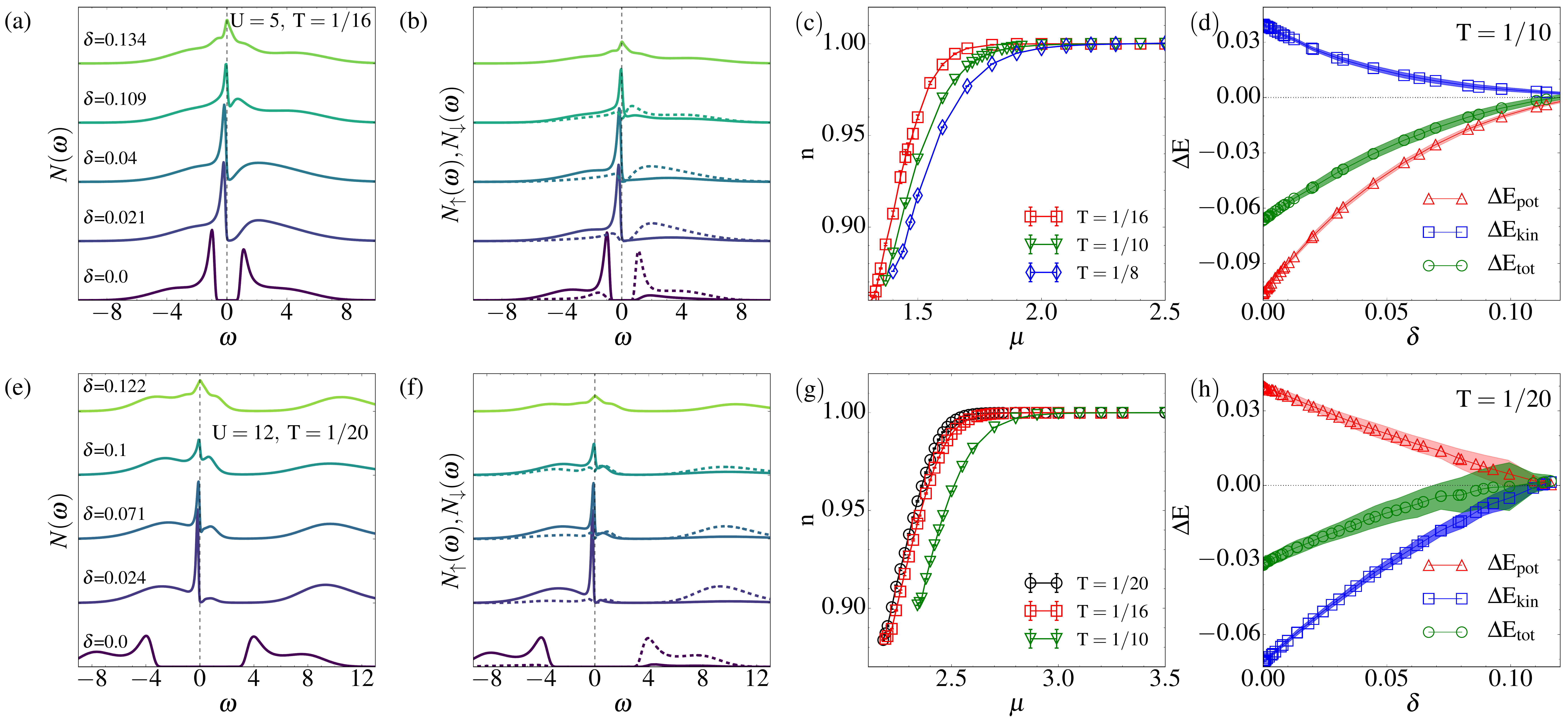}
}
\caption{Local DOS (a,e) along with spin projections (b,f) for different doping levels as obtained from analytically continued~\cite{DominicMEM} data. Occupation versus chemical potential $n(\mu)=1-\delta(\mu)$ for different temperatures (c,g). Doping dependence of the difference in potential, kinetic, and total energies between the AF and the underlying normal phase (d,h).  Data are for $t'=-0.1$ and  $U=5$ (top panels) and $U=12$ (bottom panels). For more data on $N(\omega)$ see SM~\cite{Note1} Figs.3,4.}
\label{fig3}
\end{figure*}

First, consider the $\delta=0$ case. For $t'=0$ we know that CDMFT recovers the AF insulating behavior~\cite{LorenzoAF2017}. In principle, at small $U$ and large $t'$, the AF state can have both hole- and electron pockets at the Fermi surface. Then the AF state would be metallic even at $\delta=0$~\cite{hofstetter1998, kotliarSBaf}. Here we find that the $\delta=0$ solution is insulating for all $t'$ and $U$ we considered. 
This can be checked from the local DOS and from the occupation shown at $t'=-0.1$ for $U=5$ and $U=12$ in Fig.~\ref{fig3}. More specifically, the plateau in the occupation at $n(\mu)=1$ in Figs.~\ref{fig3}c,g signals an incompressible insulator, i.e. the charge compressibility $\kappa=n^{-2} dn/d\mu$ vanishes. 

Second, consider the AF state at finite doping $\delta \neq 0$. In this case, $n(\mu)$ has a finite slope, signalling a compressible metal, i.e. $\kappa>0$. In addition, Figs.~\ref{fig3}a,e show that the local DOS has a small but finite spectral weight at the Fermi level, indicating a metallic state.  

Therefore, at $\delta=0$ there is an AF insulator (AF-I), whereas at $\delta>0$ there is an AF metal (AF-M). This also holds in the $d \rightarrow \infty$ limit~\cite{albertoAF}. What is the nature of the AF-I to AF-M transition driven by doping? Close to $n(\mu)=1$ ($\delta=1-n=0$), the occupation $n(\mu)$ is continuous for all $T$ we have explored. As $T$ decreases, the curvature at the transition becomes sharper, suggesting a discontinuous change in slope in the $T=0$ limit, as expected for a second-order AF-I to AF-M transition. The transition is a crossover at finite $T$. 

{\it Density of states and ``in-gap'' states. --} There are striking differences between the DOS $N(\omega)$ of a doped Slater AF ($U=5$) {\it vs} a doped Mott AF ($U=12$). 

For $U=5$, the $N(\omega)$ spectra for $\delta=0$ in Fig.~\ref{fig3}a shows two Bogoliubov peaks along with high frequency precursors of the Hubbard bands~\cite{Moreo1995}. 
When $\mu$ reaches the edge of the lower Bogoliubov peak, metallic behavior is recoverd since doped holes appear at $\omega=0$. The rearrangement of the specral weight is not expected from the HF Slater solution. Upon doping, spectral weight transfers from high to low frequencies: the lower Bogoliubov peak decreases in intensity and moves close to the Fermi energy $\omega=0$ and, correspondingly, the upper Bogoliubov peak broadens. 
Fig.~\ref{fig3}b shows that the upper and lower Bogoliubov peaks have sizeable spin polarization, as in the $t'=0$ case~\cite{LorenzoAF2017}. 
 
By contrast, for $U=12$, the spectra for $\delta=0$ in Fig.~\ref{fig3}c,d have a clear four peak structure: two Bogoliubov peaks surrounded by Hubbard bands~\cite{Vilk:1997, Moreo1995, Preuss1995, LorenzoAF2017}. In the doped case, there is a dramatic redistribution of spectral weight over a large frequency range across the AF gap, reminiscent of the Eskes-Meinders-Sawatzky picture~\cite{eskes1991}: the lower Bogoliubov peak sharpens and a new spectral feature, ``in-gap state'', appears between the upper Hubbard band and the Fermi level located at $\omega=0$. In that picture, the lower Hubbard band comes mostly from removing electrons in singly-occupied sites, while the upper Hubbard band comes mostly from adding electrons to singly-occupied sites. Given the large local moment at this value of $U$ in the AF, this is consistent with the fact that these Hubbard bands are strongly spin polarized, as seen in Fig.~\ref{fig3}f.  On the other hand, the ``in-gap states'' come, in that same picture~\cite{eskes1991}, from adding electrons in empty sites, which explains the near absence of spin polarization observed in Fig.~\ref{fig3}f. Finally, with further doping, the lower Bogoliubov peak and the Hubbard bands decrease in intensity at the expense of the in-gap state above $\omega=0$. These results are compatible with the variational approach in Ref.~\onlinecite{wuAFevolution}.

{\it Stability of the N\'eel phase. --} 
To assess the origin of the stability of the AF phase, we compare the kinetic, potential and total energy differences between the AF phase and the underlying normal phase~\cite{TarantoPRB2012,Tocchio2016,LorenzoAF2017} as a function of doping $\delta$ at low $T$. Fig.~\ref{fig3}d,h shows that the crossover in the source of stability of the AF phase that we identified earlier~\cite{LorenzoAF2017} at $\delta=0$ persists for all doping levels. Therefore the hidden Mott transition at $\delta=0$ reorganises the energetics both of the AF-I at $\delta=0$ and of the AF-M away from $\delta=0$.

{\it Discussion. --} 
Our results are relevant for experiments with ultracold atoms and with cuprates. The first observation of the AF phase in a 2D square optical lattice appeared recently~\cite{mazurenko2017cold}. Persistence of AF correlations was found up to $\delta \approx 0.15$ for $U=7.2$ and $t'=0$. Consistency of this finding with our results is promising.
Our $U-\delta-T$ map of the AF phase can be explored further with ultracold atom systems since $U$, $\delta$ and $T$ can be tuned. 
Specifically the nonmonotonic behavior of $\delta_N^d(U)$, along with stronger reduction of $T^d_N(U)$ with increasing $|t'|$ at small $U$, are testable predictions. 
 
When comparing our $U-T-\delta$ map with experiments on hole-doped cuprates, one should focus on the multilayer case where interlayer magnetic exchange favors mean-field like behavior. In the $n=5$ CuO$_2$ layer cuprates, AF persists up to $\delta \approx 0.10$ and it decreases with decreasing $n$~\cite{mukuda2008, mukuda2012}. Our predictions for $m_z$ could be compared with neutron scattering and muon spin rotation experiments in this regime. 
Other effects that we did not take into account and that can decrease $\delta_{\rm N}^d(U)$ are the development of incommensurate spin-density waves and competition with other phases. 

The $n(\mu)$ curve and the resulting charge compressibility $\kappa$ that describes the continuous transition between an AF insulator and an AF metal as a function of doping is another prediction that can be tested with ultracold atoms. In principle, such measurements are also possible in cuprates~\cite{InoMU, HarimaMU,RietveldMU, dirkMU}. 

The in-gap state feature that we found in the DOS has a position and a width that is compatible with recent scanning tunnelling microscopy experiments on lightly doped AF Mott insulators~\cite{CaiSTM, YeSTM}. In addition, the observed transfer of spectral weight from high energy to low energy as a function of doping is consistent with our results. We predict that a spin-polarized STM probe will find that these states are essentially unpolarized, by contrast with the lower and upper Hubbard bands~\footnote{In these experiments, the DOS at $\omega=0$ vanishes while it in our case it is small. The inhomogeneity of the samples suggests that disorder can induce localisation effects at $\omega=0$, as pointed out in Ref.~\onlinecite{wuAFevolution}. The AF phase of cuprates at finite doping is generally considered metallic~\protect\cite{ando2001}.}. 

Our prediction that the doped AF state is stabilized by a gain in kinetic energy for large $U$ and by a gain in potential energy for small $U$ can in principle be tested by optical spectroscopy in cuprates~\cite{Molegraaf2002,deutscher2005,PhysRevB.74.064510}. If the correlation strength $U$ is lower in electron that in hole doped cuprates, as has been proposed~\cite{st,Weber:2010}, our data suggest a potential energy driven AF in electron doped cuprates and a kinetic energy driven AF in hole doped cuprates, similar to earlier findings on the Emery model~\cite{cedricApical}. 

Moving forwards, it will be important to study the interplay between antiferromagnetism, pseudogap and superconductivity~\cite{sht, sht2, ssht, sshtRHO, LorenzoSC, Lorenzo3band}.

\begin{acknowledgments}
We acknowledge discussion with D. S\'en\'echal and K. Miyake. This work has been supported by the Natural Sciences and Engineering Research Council of Canada (NSERC) under grant RGPIN-2014-04584, the Canada First Research Excellence Fund and by the Research Chair in the Theory of Quantum Materials. Simulations were performed on computers provided by the Canadian Foundation for Innovation, the Minist\`ere de l'\'Education des Loisirs et du Sport (Qu\'ebec), Calcul Qu\'ebec, and Compute Canada.
\end{acknowledgments}


\begin{thebibliography}{80}%
\makeatletter
\providecommand \@ifxundefined [1]{%
 \@ifx{#1\undefined}
}%
\providecommand \@ifnum [1]{%
 \ifnum #1\expandafter \@firstoftwo
 \else \expandafter \@secondoftwo
 \fi
}%
\providecommand \@ifx [1]{%
 \ifx #1\expandafter \@firstoftwo
 \else \expandafter \@secondoftwo
 \fi
}%
\providecommand \natexlab [1]{#1}%
\providecommand \enquote  [1]{``#1''}%
\providecommand \bibnamefont  [1]{#1}%
\providecommand \bibfnamefont [1]{#1}%
\providecommand \citenamefont [1]{#1}%
\providecommand \href@noop [0]{\@secondoftwo}%
\providecommand \href [0]{\begingroup \@sanitize@url \@href}%
\providecommand \@href[1]{\@@startlink{#1}\@@href}%
\providecommand \@@href[1]{\endgroup#1\@@endlink}%
\providecommand \@sanitize@url [0]{\catcode `\\12\catcode `\$12\catcode
  `\&12\catcode `\#12\catcode `\^12\catcode `\_12\catcode `\%12\relax}%
\providecommand \@@startlink[1]{}%
\providecommand \@@endlink[0]{}%
\providecommand \url  [0]{\begingroup\@sanitize@url \@url }%
\providecommand \@url [1]{\endgroup\@href {#1}{\urlprefix }}%
\providecommand \urlprefix  [0]{URL }%
\providecommand \Eprint [0]{\href }%
\providecommand \doibase [0]{http://dx.doi.org/}%
\providecommand \selectlanguage [0]{\@gobble}%
\providecommand \bibinfo  [0]{\@secondoftwo}%
\providecommand \bibfield  [0]{\@secondoftwo}%
\providecommand \translation [1]{[#1]}%
\providecommand \BibitemOpen [0]{}%
\providecommand \bibitemStop [0]{}%
\providecommand \bibitemNoStop [0]{.\EOS\space}%
\providecommand \EOS [0]{\spacefactor3000\relax}%
\providecommand \BibitemShut  [1]{\csname bibitem#1\endcsname}%
\let\auto@bib@innerbib\@empty
\bibitem [{\citenamefont {Keimer}\ \emph {et~al.}(2015)\citenamefont {Keimer},
  \citenamefont {Kivelson}, \citenamefont {Norman}, \citenamefont {Uchida},\
  and\ \citenamefont {Zaanen}}]{keimerRev}%
  \BibitemOpen
  \bibfield  {author} {\bibinfo {author} {\bibfnamefont {B.}~\bibnamefont
  {Keimer}}, \bibinfo {author} {\bibfnamefont {S.~A.}\ \bibnamefont
  {Kivelson}}, \bibinfo {author} {\bibfnamefont {M.~R.}\ \bibnamefont
  {Norman}}, \bibinfo {author} {\bibfnamefont {S.}~\bibnamefont {Uchida}}, \
  and\ \bibinfo {author} {\bibfnamefont {J.}~\bibnamefont {Zaanen}},\ }\href
  {\doibase 10.1038/nature14165} {\bibfield  {journal} {\bibinfo  {journal}
  {Nature}\ }\textbf {\bibinfo {volume} {518}},\ \bibinfo {pages} {179}
  (\bibinfo {year} {2015})}\BibitemShut {NoStop}%
\bibitem [{\citenamefont {Anderson}(1987)}]{Anderson:1987}%
  \BibitemOpen
  \bibfield  {author} {\bibinfo {author} {\bibfnamefont {P.~W.}\ \bibnamefont
  {Anderson}},\ }\href {\doibase 10.1126/science.235.4793.1196} {\bibfield
  {journal} {\bibinfo  {journal} {Science}\ }\textbf {\bibinfo {volume}
  {235}},\ \bibinfo {pages} {1196} (\bibinfo {year} {1987})}\BibitemShut
  {NoStop}%
\bibitem [{\citenamefont {Jaksch}\ and\ \citenamefont {Zoller}(2005)}]{jzHM}%
  \BibitemOpen
  \bibfield  {author} {\bibinfo {author} {\bibfnamefont {D.}~\bibnamefont
  {Jaksch}}\ and\ \bibinfo {author} {\bibfnamefont {P.}~\bibnamefont
  {Zoller}},\ }\href@noop {} {\bibfield  {journal} {\bibinfo  {journal} {Ann.
  Phys.}\ }\textbf {\bibinfo {volume} {315}},\ \bibinfo {eid} {52} (\bibinfo
  {year} {2005})}\BibitemShut {NoStop}%
\bibitem [{\citenamefont {Tremblay}(2013)}]{AMJulich}%
  \BibitemOpen
  \bibfield  {author} {\bibinfo {author} {\bibfnamefont {A.-M.~S.}\
  \bibnamefont {Tremblay}},\ }in\ \href
  {http://juser.fz-juelich.de/record/137827/files/FZJ-2013-04137.pdf?version=1}
  {\emph {\bibinfo {booktitle} {Emergent Phenomena in Correlated Matter
  Modeling and Simulation}}},\ Vol.~\bibinfo {volume} {3},\ \bibinfo {editor}
  {edited by\ \bibinfo {editor} {\bibfnamefont {E.}~\bibnamefont {Pavarini}},
  \bibinfo {editor} {\bibfnamefont {E.}~\bibnamefont {Koch}}, \ and\ \bibinfo
  {editor} {\bibfnamefont {U.}~\bibnamefont {Schollw\"ock}}}\ (\bibinfo
  {publisher} {Verlag des Forschungszentrum},\ \bibinfo {address} {J\"ulich},\
  \bibinfo {year} {2013})\ Chap.~\bibinfo {chapter} {10}\BibitemShut {NoStop}%
\bibitem [{\citenamefont {Georges}\ and\ \citenamefont
  {Giamarchi}(2010)}]{AntoineLesHouches}%
  \BibitemOpen
  \bibfield  {author} {\bibinfo {author} {\bibfnamefont {A.}~\bibnamefont
  {Georges}}\ and\ \bibinfo {author} {\bibfnamefont {T.}~\bibnamefont
  {Giamarchi}},\ }in\ \href@noop {} {\emph {\bibinfo {booktitle} {Many-body
  Physics with Ultracold Gases}}},\ Vol.~\bibinfo {volume} {94},\ \bibinfo
  {editor} {edited by\ \bibinfo {editor} {\bibfnamefont {C.}~\bibnamefont
  {Salomon}}, \bibinfo {editor} {\bibfnamefont {G.}~\bibnamefont
  {Shlyapnikov}}, \ and\ \bibinfo {editor} {\bibfnamefont {L.}~\bibnamefont
  {Cugliandolo}}}\ (\bibinfo  {publisher} {Oxford University Press},\ \bibinfo
  {address} {Les Houches},\ \bibinfo {year} {2010})\ Chap.~\bibinfo {chapter}
  {1}\BibitemShut {NoStop}%
\bibitem [{\citenamefont {Maier}\ \emph {et~al.}(2005)\citenamefont {Maier},
  \citenamefont {Jarrell}, \citenamefont {Pruschke},\ and\ \citenamefont
  {Hettler}}]{maier}%
  \BibitemOpen
  \bibfield  {author} {\bibinfo {author} {\bibfnamefont {T.}~\bibnamefont
  {Maier}}, \bibinfo {author} {\bibfnamefont {M.}~\bibnamefont {Jarrell}},
  \bibinfo {author} {\bibfnamefont {T.}~\bibnamefont {Pruschke}}, \ and\
  \bibinfo {author} {\bibfnamefont {M.~H.}\ \bibnamefont {Hettler}},\ }\href
  {\doibase 10.1103/RevModPhys.77.1027} {\bibfield  {journal} {\bibinfo
  {journal} {Rev. Mod. Phys.}\ }\textbf {\bibinfo {volume} {77}},\ \bibinfo
  {pages} {1027} (\bibinfo {year} {2005})}\BibitemShut {NoStop}%
\bibitem [{\citenamefont {Kotliar}\ \emph {et~al.}(2006)\citenamefont
  {Kotliar}, \citenamefont {Savrasov}, \citenamefont {Haule}, \citenamefont
  {Oudovenko}, \citenamefont {Parcollet},\ and\ \citenamefont
  {Marianetti}}]{kotliarRMP}%
  \BibitemOpen
  \bibfield  {author} {\bibinfo {author} {\bibfnamefont {G.}~\bibnamefont
  {Kotliar}}, \bibinfo {author} {\bibfnamefont {S.~Y.}\ \bibnamefont
  {Savrasov}}, \bibinfo {author} {\bibfnamefont {K.}~\bibnamefont {Haule}},
  \bibinfo {author} {\bibfnamefont {V.~S.}\ \bibnamefont {Oudovenko}}, \bibinfo
  {author} {\bibfnamefont {O.}~\bibnamefont {Parcollet}}, \ and\ \bibinfo
  {author} {\bibfnamefont {C.~A.}\ \bibnamefont {Marianetti}},\ }\href
  {\doibase 10.1103/RevModPhys.78.865} {\bibfield  {journal} {\bibinfo
  {journal} {Rev. Mod. Phys.}\ }\textbf {\bibinfo {volume} {78}},\ \bibinfo
  {eid} {865} (\bibinfo {year} {2006})}\BibitemShut {NoStop}%
\bibitem [{\citenamefont {Tremblay}\ \emph {et~al.}(2006)\citenamefont
  {Tremblay}, \citenamefont {Kyung},\ and\ \citenamefont
  {S\'{e}n\'{e}chal}}]{tremblayR}%
  \BibitemOpen
  \bibfield  {author} {\bibinfo {author} {\bibfnamefont {A.-M.~S.}\
  \bibnamefont {Tremblay}}, \bibinfo {author} {\bibfnamefont {B.}~\bibnamefont
  {Kyung}}, \ and\ \bibinfo {author} {\bibfnamefont {D.}~\bibnamefont
  {S\'{e}n\'{e}chal}},\ }\href {\doibase 10.1063/1.2199446} {\bibfield
  {journal} {\bibinfo  {journal} {Low Temp. Phys.}\ }\textbf {\bibinfo {volume}
  {32}},\ \bibinfo {pages} {424} (\bibinfo {year} {2006})}\BibitemShut
  {NoStop}%
\bibitem [{\citenamefont {Greif}\ \emph {et~al.}(2013)\citenamefont {Greif},
  \citenamefont {Uehlinger}, \citenamefont {Jotzu}, \citenamefont {Tarruell},\
  and\ \citenamefont {Esslinger}}]{Greif:2013}%
  \BibitemOpen
  \bibfield  {author} {\bibinfo {author} {\bibfnamefont {D.}~\bibnamefont
  {Greif}}, \bibinfo {author} {\bibfnamefont {T.}~\bibnamefont {Uehlinger}},
  \bibinfo {author} {\bibfnamefont {G.}~\bibnamefont {Jotzu}}, \bibinfo
  {author} {\bibfnamefont {L.}~\bibnamefont {Tarruell}}, \ and\ \bibinfo
  {author} {\bibfnamefont {T.}~\bibnamefont {Esslinger}},\ }\href {\doibase
  10.1126/science.1236362} {\bibfield  {journal} {\bibinfo  {journal}
  {Science}\ }\textbf {\bibinfo {volume} {340}},\ \bibinfo {pages} {1307}
  (\bibinfo {year} {2013})}\BibitemShut {NoStop}%
\bibitem [{\citenamefont {{Hart}}\ \emph {et~al.}(2015)\citenamefont {{Hart}},
  \citenamefont {{Duarte}}, \citenamefont {{Yang}}, \citenamefont {{Liu}},
  \citenamefont {{Paiva}}, \citenamefont {{Khatami}}, \citenamefont
  {{Scalettar}}, \citenamefont {{Trivedi}}, \citenamefont {{Huse}},\ and\
  \citenamefont {{Hulet}}}]{Hart:2015}%
  \BibitemOpen
  \bibfield  {author} {\bibinfo {author} {\bibfnamefont {R.~A.}\ \bibnamefont
  {{Hart}}}, \bibinfo {author} {\bibfnamefont {P.~M.}\ \bibnamefont
  {{Duarte}}}, \bibinfo {author} {\bibfnamefont {T.-L.}\ \bibnamefont
  {{Yang}}}, \bibinfo {author} {\bibfnamefont {X.}~\bibnamefont {{Liu}}},
  \bibinfo {author} {\bibfnamefont {T.}~\bibnamefont {{Paiva}}}, \bibinfo
  {author} {\bibfnamefont {E.}~\bibnamefont {{Khatami}}}, \bibinfo {author}
  {\bibfnamefont {R.~T.}\ \bibnamefont {{Scalettar}}}, \bibinfo {author}
  {\bibfnamefont {N.}~\bibnamefont {{Trivedi}}}, \bibinfo {author}
  {\bibfnamefont {D.~A.}\ \bibnamefont {{Huse}}}, \ and\ \bibinfo {author}
  {\bibfnamefont {R.~G.}\ \bibnamefont {{Hulet}}},\ }\href {\doibase
  10.1038/nature14223} {\bibfield  {journal} {\bibinfo  {journal} {Nature}\
  }\textbf {\bibinfo {volume} {519}},\ \bibinfo {pages} {211} (\bibinfo {year}
  {2015})}\BibitemShut {NoStop}%
\bibitem [{\citenamefont {Parsons}\ \emph {et~al.}(2016)\citenamefont
  {Parsons}, \citenamefont {Mazurenko}, \citenamefont {Chiu}, \citenamefont
  {Ji}, \citenamefont {Greif},\ and\ \citenamefont {Greiner}}]{Parsons:2016}%
  \BibitemOpen
  \bibfield  {author} {\bibinfo {author} {\bibfnamefont {M.~F.}\ \bibnamefont
  {Parsons}}, \bibinfo {author} {\bibfnamefont {A.}~\bibnamefont {Mazurenko}},
  \bibinfo {author} {\bibfnamefont {C.~S.}\ \bibnamefont {Chiu}}, \bibinfo
  {author} {\bibfnamefont {G.}~\bibnamefont {Ji}}, \bibinfo {author}
  {\bibfnamefont {D.}~\bibnamefont {Greif}}, \ and\ \bibinfo {author}
  {\bibfnamefont {M.}~\bibnamefont {Greiner}},\ }\href {\doibase
  10.1126/science.aag1430} {\bibfield  {journal} {\bibinfo  {journal}
  {Science}\ }\textbf {\bibinfo {volume} {353}},\ \bibinfo {pages} {1253}
  (\bibinfo {year} {2016})}\BibitemShut {NoStop}%
\bibitem [{\citenamefont {Boll}\ \emph {et~al.}(2016)\citenamefont {Boll},
  \citenamefont {Hilker}, \citenamefont {Salomon}, \citenamefont {Omran},
  \citenamefont {Nespolo}, \citenamefont {Pollet}, \citenamefont {Bloch},\ and\
  \citenamefont {Gross}}]{Boll:2016}%
  \BibitemOpen
  \bibfield  {author} {\bibinfo {author} {\bibfnamefont {M.}~\bibnamefont
  {Boll}}, \bibinfo {author} {\bibfnamefont {T.~A.}\ \bibnamefont {Hilker}},
  \bibinfo {author} {\bibfnamefont {G.}~\bibnamefont {Salomon}}, \bibinfo
  {author} {\bibfnamefont {A.}~\bibnamefont {Omran}}, \bibinfo {author}
  {\bibfnamefont {J.}~\bibnamefont {Nespolo}}, \bibinfo {author} {\bibfnamefont
  {L.}~\bibnamefont {Pollet}}, \bibinfo {author} {\bibfnamefont
  {I.}~\bibnamefont {Bloch}}, \ and\ \bibinfo {author} {\bibfnamefont
  {C.}~\bibnamefont {Gross}},\ }\href {\doibase 10.1126/science.aag1635}
  {\bibfield  {journal} {\bibinfo  {journal} {Science}\ }\textbf {\bibinfo
  {volume} {353}},\ \bibinfo {pages} {1257} (\bibinfo {year}
  {2016})}\BibitemShut {NoStop}%
\bibitem [{\citenamefont {Cheuk}\ \emph {et~al.}(2016)\citenamefont {Cheuk},
  \citenamefont {Nichols}, \citenamefont {Lawrence}, \citenamefont {Okan},
  \citenamefont {Zhang}, \citenamefont {Khatami}, \citenamefont {Trivedi},
  \citenamefont {Paiva}, \citenamefont {Rigol},\ and\ \citenamefont
  {Zwierlein}}]{Cheuk:2016}%
  \BibitemOpen
  \bibfield  {author} {\bibinfo {author} {\bibfnamefont {L.~W.}\ \bibnamefont
  {Cheuk}}, \bibinfo {author} {\bibfnamefont {M.~A.}\ \bibnamefont {Nichols}},
  \bibinfo {author} {\bibfnamefont {K.~R.}\ \bibnamefont {Lawrence}}, \bibinfo
  {author} {\bibfnamefont {M.}~\bibnamefont {Okan}}, \bibinfo {author}
  {\bibfnamefont {H.}~\bibnamefont {Zhang}}, \bibinfo {author} {\bibfnamefont
  {E.}~\bibnamefont {Khatami}}, \bibinfo {author} {\bibfnamefont
  {N.}~\bibnamefont {Trivedi}}, \bibinfo {author} {\bibfnamefont
  {T.}~\bibnamefont {Paiva}}, \bibinfo {author} {\bibfnamefont
  {M.}~\bibnamefont {Rigol}}, \ and\ \bibinfo {author} {\bibfnamefont {M.~W.}\
  \bibnamefont {Zwierlein}},\ }\href {\doibase 10.1126/science.aag3349}
  {\bibfield  {journal} {\bibinfo  {journal} {Science}\ }\textbf {\bibinfo
  {volume} {353}},\ \bibinfo {pages} {1260} (\bibinfo {year}
  {2016})}\BibitemShut {NoStop}%
\bibitem [{\citenamefont {Mazurenko}\ \emph {et~al.}(2017)\citenamefont
  {Mazurenko}, \citenamefont {Chiu}, \citenamefont {Ji}, \citenamefont
  {Parsons}, \citenamefont {Kan{\'a}sz-Nagy}, \citenamefont {Schmidt},
  \citenamefont {Grusdt}, \citenamefont {Demler}, \citenamefont {Greif},\ and\
  \citenamefont {Greiner}}]{mazurenko2017cold}%
  \BibitemOpen
  \bibfield  {author} {\bibinfo {author} {\bibfnamefont {A.}~\bibnamefont
  {Mazurenko}}, \bibinfo {author} {\bibfnamefont {C.~S.}\ \bibnamefont {Chiu}},
  \bibinfo {author} {\bibfnamefont {G.}~\bibnamefont {Ji}}, \bibinfo {author}
  {\bibfnamefont {M.~F.}\ \bibnamefont {Parsons}}, \bibinfo {author}
  {\bibfnamefont {M.}~\bibnamefont {Kan{\'a}sz-Nagy}}, \bibinfo {author}
  {\bibfnamefont {R.}~\bibnamefont {Schmidt}}, \bibinfo {author} {\bibfnamefont
  {F.}~\bibnamefont {Grusdt}}, \bibinfo {author} {\bibfnamefont
  {E.}~\bibnamefont {Demler}}, \bibinfo {author} {\bibfnamefont
  {D.}~\bibnamefont {Greif}}, \ and\ \bibinfo {author} {\bibfnamefont
  {M.}~\bibnamefont {Greiner}},\ }\href@noop {} {\bibfield  {journal} {\bibinfo
   {journal} {Nature}\ }\textbf {\bibinfo {volume} {545}},\ \bibinfo {pages}
  {462} (\bibinfo {year} {2017})}\BibitemShut {NoStop}%
\bibitem [{\citenamefont {Drewes}\ \emph {et~al.}(2017)\citenamefont {Drewes},
  \citenamefont {Miller}, \citenamefont {Cocchi}, \citenamefont {Chan},
  \citenamefont {Wurz}, \citenamefont {Gall}, \citenamefont {Pertot},
  \citenamefont {Brennecke},\ and\ \citenamefont {K\"ohl}}]{drewesPRL2017}%
  \BibitemOpen
  \bibfield  {author} {\bibinfo {author} {\bibfnamefont {J.~H.}\ \bibnamefont
  {Drewes}}, \bibinfo {author} {\bibfnamefont {L.~A.}\ \bibnamefont {Miller}},
  \bibinfo {author} {\bibfnamefont {E.}~\bibnamefont {Cocchi}}, \bibinfo
  {author} {\bibfnamefont {C.~F.}\ \bibnamefont {Chan}}, \bibinfo {author}
  {\bibfnamefont {N.}~\bibnamefont {Wurz}}, \bibinfo {author} {\bibfnamefont
  {M.}~\bibnamefont {Gall}}, \bibinfo {author} {\bibfnamefont {D.}~\bibnamefont
  {Pertot}}, \bibinfo {author} {\bibfnamefont {F.}~\bibnamefont {Brennecke}}, \
  and\ \bibinfo {author} {\bibfnamefont {M.}~\bibnamefont {K\"ohl}},\ }\href
  {\doibase 10.1103/PhysRevLett.118.170401} {\bibfield  {journal} {\bibinfo
  {journal} {Phys. Rev. Lett.}\ }\textbf {\bibinfo {volume} {118}},\ \bibinfo
  {pages} {170401} (\bibinfo {year} {2017})}\BibitemShut {NoStop}%
\bibitem [{\citenamefont {{Cai}}\ \emph {et~al.}(2016)\citenamefont {{Cai}},
  \citenamefont {{Ruan}}, \citenamefont {{Peng}}, \citenamefont {{Ye}},
  \citenamefont {{Li}}, \citenamefont {{Hao}}, \citenamefont {{Zhou}},
  \citenamefont {{Lee}},\ and\ \citenamefont {{Wang}}}]{CaiSTM}%
  \BibitemOpen
  \bibfield  {author} {\bibinfo {author} {\bibfnamefont {P.}~\bibnamefont
  {{Cai}}}, \bibinfo {author} {\bibfnamefont {W.}~\bibnamefont {{Ruan}}},
  \bibinfo {author} {\bibfnamefont {Y.}~\bibnamefont {{Peng}}}, \bibinfo
  {author} {\bibfnamefont {C.}~\bibnamefont {{Ye}}}, \bibinfo {author}
  {\bibfnamefont {X.}~\bibnamefont {{Li}}}, \bibinfo {author} {\bibfnamefont
  {Z.}~\bibnamefont {{Hao}}}, \bibinfo {author} {\bibfnamefont
  {X.}~\bibnamefont {{Zhou}}}, \bibinfo {author} {\bibfnamefont {D.-H.}\
  \bibnamefont {{Lee}}}, \ and\ \bibinfo {author} {\bibfnamefont
  {Y.}~\bibnamefont {{Wang}}},\ }\href {\doibase 10.1038/nphys3840} {\bibfield
  {journal} {\bibinfo  {journal} {Nature Physics}\ }\textbf {\bibinfo {volume}
  {12}},\ \bibinfo {pages} {1047} (\bibinfo {year} {2016})},\ \Eprint
  {http://arxiv.org/abs/1508.05586} {arXiv:1508.05586 [cond-mat.supr-con]}
  \BibitemShut {NoStop}%
\bibitem [{\citenamefont {{Ye}}\ \emph {et~al.}(2013)\citenamefont {{Ye}},
  \citenamefont {{Cai}}, \citenamefont {{Yu}}, \citenamefont {{Zhou}},
  \citenamefont {{Ruan}}, \citenamefont {{Liu}}, \citenamefont {{Jin}},\ and\
  \citenamefont {{Wang}}}]{YeSTM}%
  \BibitemOpen
  \bibfield  {author} {\bibinfo {author} {\bibfnamefont {C.}~\bibnamefont
  {{Ye}}}, \bibinfo {author} {\bibfnamefont {P.}~\bibnamefont {{Cai}}},
  \bibinfo {author} {\bibfnamefont {R.}~\bibnamefont {{Yu}}}, \bibinfo {author}
  {\bibfnamefont {X.}~\bibnamefont {{Zhou}}}, \bibinfo {author} {\bibfnamefont
  {W.}~\bibnamefont {{Ruan}}}, \bibinfo {author} {\bibfnamefont
  {Q.}~\bibnamefont {{Liu}}}, \bibinfo {author} {\bibfnamefont
  {C.}~\bibnamefont {{Jin}}}, \ and\ \bibinfo {author} {\bibfnamefont
  {Y.}~\bibnamefont {{Wang}}},\ }\href {\doibase 10.1038/ncomms2369} {\bibfield
   {journal} {\bibinfo  {journal} {Nature Communications}\ }\textbf {\bibinfo
  {volume} {4}},\ \bibinfo {eid} {1365} (\bibinfo {year} {2013})},\ \Eprint
  {http://arxiv.org/abs/1201.0342} {arXiv:1201.0342 [cond-mat.str-el]}
  \BibitemShut {NoStop}%
\bibitem [{\citenamefont {Borejsza}\ and\ \citenamefont
  {Dupuis}(2004)}]{Dupuis2004}%
  \BibitemOpen
  \bibfield  {author} {\bibinfo {author} {\bibfnamefont {K.}~\bibnamefont
  {Borejsza}}\ and\ \bibinfo {author} {\bibfnamefont {N.}~\bibnamefont
  {Dupuis}},\ }\href {\doibase 10.1103/PhysRevB.69.085119} {\bibfield
  {journal} {\bibinfo  {journal} {Phys. Rev. B}\ }\textbf {\bibinfo {volume}
  {69}},\ \bibinfo {pages} {085119} (\bibinfo {year} {2004})}\BibitemShut
  {NoStop}%
\bibitem [{\citenamefont {S\'en\'echal}\ \emph {et~al.}(2005)\citenamefont
  {S\'en\'echal}, \citenamefont {Lavertu}, \citenamefont {Marois},\ and\
  \citenamefont {Tremblay}}]{senechalAFSC2005}%
  \BibitemOpen
  \bibfield  {author} {\bibinfo {author} {\bibfnamefont {D.}~\bibnamefont
  {S\'en\'echal}}, \bibinfo {author} {\bibfnamefont {P.-L.}\ \bibnamefont
  {Lavertu}}, \bibinfo {author} {\bibfnamefont {M.-A.}\ \bibnamefont {Marois}},
  \ and\ \bibinfo {author} {\bibfnamefont {A.-M.~S.}\ \bibnamefont
  {Tremblay}},\ }\href {\doibase 10.1103/PhysRevLett.94.156404} {\bibfield
  {journal} {\bibinfo  {journal} {Phys. Rev. Lett.}\ }\textbf {\bibinfo
  {volume} {94}},\ \bibinfo {pages} {156404} (\bibinfo {year}
  {2005})}\BibitemShut {NoStop}%
\bibitem [{\citenamefont {{Aichhorn, M.}}\ and\ \citenamefont {{Arrigoni,
  E.}}(2005)}]{markus2005}%
  \BibitemOpen
  \bibfield  {author} {\bibinfo {author} {\bibnamefont {{Aichhorn, M.}}}\ and\
  \bibinfo {author} {\bibnamefont {{Arrigoni, E.}}},\ }\href {\doibase
  10.1209/epl/i2005-10192-1} {\bibfield  {journal} {\bibinfo  {journal}
  {Europhys. Lett.}\ }\textbf {\bibinfo {volume} {72}},\ \bibinfo {pages} {117}
  (\bibinfo {year} {2005})}\BibitemShut {NoStop}%
\bibitem [{\citenamefont {Aichhorn}\ \emph {et~al.}(2007)\citenamefont
  {Aichhorn}, \citenamefont {Arrigoni}, \citenamefont {Potthoff},\ and\
  \citenamefont {Hanke}}]{markus}%
  \BibitemOpen
  \bibfield  {author} {\bibinfo {author} {\bibfnamefont {M.}~\bibnamefont
  {Aichhorn}}, \bibinfo {author} {\bibfnamefont {E.}~\bibnamefont {Arrigoni}},
  \bibinfo {author} {\bibfnamefont {M.}~\bibnamefont {Potthoff}}, \ and\
  \bibinfo {author} {\bibfnamefont {W.}~\bibnamefont {Hanke}},\ }\href
  {\doibase 10.1103/PhysRevB.76.224509} {\bibfield  {journal} {\bibinfo
  {journal} {Phys. Rev. B}\ }\textbf {\bibinfo {volume} {76}},\ \bibinfo
  {pages} {224509} (\bibinfo {year} {2007})}\BibitemShut {NoStop}%
\bibitem [{\citenamefont {Tocchio}\ \emph {et~al.}(2016)\citenamefont
  {Tocchio}, \citenamefont {Becca},\ and\ \citenamefont
  {Sorella}}]{Tocchio2016}%
  \BibitemOpen
  \bibfield  {author} {\bibinfo {author} {\bibfnamefont {L.~F.}\ \bibnamefont
  {Tocchio}}, \bibinfo {author} {\bibfnamefont {F.}~\bibnamefont {Becca}}, \
  and\ \bibinfo {author} {\bibfnamefont {S.}~\bibnamefont {Sorella}},\ }\href
  {\doibase 10.1103/PhysRevB.94.195126} {\bibfield  {journal} {\bibinfo
  {journal} {Phys. Rev. B}\ }\textbf {\bibinfo {volume} {94}},\ \bibinfo
  {pages} {195126} (\bibinfo {year} {2016})}\BibitemShut {NoStop}%
\bibitem [{\citenamefont {Zheng}\ and\ \citenamefont {Chan}(2016)}]{ZhengDMET}%
  \BibitemOpen
  \bibfield  {author} {\bibinfo {author} {\bibfnamefont {B.-X.}\ \bibnamefont
  {Zheng}}\ and\ \bibinfo {author} {\bibfnamefont {G.~K.-L.}\ \bibnamefont
  {Chan}},\ }\href {\doibase 10.1103/PhysRevB.93.035126} {\bibfield  {journal}
  {\bibinfo  {journal} {Phys. Rev. B}\ }\textbf {\bibinfo {volume} {93}},\
  \bibinfo {pages} {035126} (\bibinfo {year} {2016})}\BibitemShut {NoStop}%
\bibitem [{\citenamefont {Hirsch}(1985)}]{Hirsch:1985}%
  \BibitemOpen
  \bibfield  {author} {\bibinfo {author} {\bibfnamefont {J.~E.}\ \bibnamefont
  {Hirsch}},\ }\href {\doibase 10.1103/PhysRevB.31.4403} {\bibfield  {journal}
  {\bibinfo  {journal} {Phys. Rev. B}\ }\textbf {\bibinfo {volume} {31}},\
  \bibinfo {pages} {4403} (\bibinfo {year} {1985})}\BibitemShut {NoStop}%
\bibitem [{\citenamefont {White}\ \emph {et~al.}(1989)\citenamefont {White},
  \citenamefont {Scalapino}, \citenamefont {Sugar}, \citenamefont {Loh},
  \citenamefont {Gubernatis},\ and\ \citenamefont {Scalettar}}]{White:1989}%
  \BibitemOpen
  \bibfield  {author} {\bibinfo {author} {\bibfnamefont {S.~R.}\ \bibnamefont
  {White}}, \bibinfo {author} {\bibfnamefont {D.~J.}\ \bibnamefont
  {Scalapino}}, \bibinfo {author} {\bibfnamefont {R.~L.}\ \bibnamefont
  {Sugar}}, \bibinfo {author} {\bibfnamefont {E.~Y.}\ \bibnamefont {Loh}},
  \bibinfo {author} {\bibfnamefont {J.~E.}\ \bibnamefont {Gubernatis}}, \ and\
  \bibinfo {author} {\bibfnamefont {R.~T.}\ \bibnamefont {Scalettar}},\ }\href
  {\doibase 10.1103/PhysRevB.40.506} {\bibfield  {journal} {\bibinfo  {journal}
  {Phys. Rev. B}\ }\textbf {\bibinfo {volume} {40}},\ \bibinfo {pages} {506}
  (\bibinfo {year} {1989})}\BibitemShut {NoStop}%
\bibitem [{\citenamefont {Lichtenstein}\ and\ \citenamefont
  {Katsnelson}(2000)}]{lkAF}%
  \BibitemOpen
  \bibfield  {author} {\bibinfo {author} {\bibfnamefont {A.~I.}\ \bibnamefont
  {Lichtenstein}}\ and\ \bibinfo {author} {\bibfnamefont {M.~I.}\ \bibnamefont
  {Katsnelson}},\ }\href {\doibase 10.1103/PhysRevB.62.R9283} {\bibfield
  {journal} {\bibinfo  {journal} {Phys. Rev. B}\ }\textbf {\bibinfo {volume}
  {62}},\ \bibinfo {pages} {R9283} (\bibinfo {year} {2000})}\BibitemShut
  {NoStop}%
\bibitem [{\citenamefont {Paiva}\ \emph {et~al.}(2001)\citenamefont {Paiva},
  \citenamefont {Scalettar}, \citenamefont {Huscroft},\ and\ \citenamefont
  {McMahan}}]{Paiva:2001}%
  \BibitemOpen
  \bibfield  {author} {\bibinfo {author} {\bibfnamefont {T.}~\bibnamefont
  {Paiva}}, \bibinfo {author} {\bibfnamefont {R.~T.}\ \bibnamefont
  {Scalettar}}, \bibinfo {author} {\bibfnamefont {C.}~\bibnamefont {Huscroft}},
  \ and\ \bibinfo {author} {\bibfnamefont {A.~K.}\ \bibnamefont {McMahan}},\
  }\href {\doibase 10.1103/PhysRevB.63.125116} {\bibfield  {journal} {\bibinfo
  {journal} {Phys. Rev. B}\ }\textbf {\bibinfo {volume} {63}},\ \bibinfo
  {pages} {125116} (\bibinfo {year} {2001})}\BibitemShut {NoStop}%
\bibitem [{\citenamefont {Kyung}\ \emph {et~al.}(2006)\citenamefont {Kyung},
  \citenamefont {Kancharla}, \citenamefont {S\'{e}n\'{e}chal}, \citenamefont
  {Tremblay}, \citenamefont {Civelli},\ and\ \citenamefont {Kotliar}}]{kyung}%
  \BibitemOpen
  \bibfield  {author} {\bibinfo {author} {\bibfnamefont {B.}~\bibnamefont
  {Kyung}}, \bibinfo {author} {\bibfnamefont {S.~S.}\ \bibnamefont
  {Kancharla}}, \bibinfo {author} {\bibfnamefont {D.}~\bibnamefont
  {S\'{e}n\'{e}chal}}, \bibinfo {author} {\bibfnamefont {A.-M.~S.}\
  \bibnamefont {Tremblay}}, \bibinfo {author} {\bibfnamefont {M.}~\bibnamefont
  {Civelli}}, \ and\ \bibinfo {author} {\bibfnamefont {G.}~\bibnamefont
  {Kotliar}},\ }\href {\doibase 10.1103/PhysRevB.73.165114} {\bibfield
  {journal} {\bibinfo  {journal} {Phys. Rev. B}\ }\textbf {\bibinfo {volume}
  {73}},\ \bibinfo {eid} {165114} (\bibinfo {year} {2006})}\BibitemShut
  {NoStop}%
\bibitem [{\citenamefont {Paiva}\ \emph {et~al.}(2010)\citenamefont {Paiva},
  \citenamefont {Scalettar}, \citenamefont {Randeria},\ and\ \citenamefont
  {Trivedi}}]{Paiva2010}%
  \BibitemOpen
  \bibfield  {author} {\bibinfo {author} {\bibfnamefont {T.}~\bibnamefont
  {Paiva}}, \bibinfo {author} {\bibfnamefont {R.}~\bibnamefont {Scalettar}},
  \bibinfo {author} {\bibfnamefont {M.}~\bibnamefont {Randeria}}, \ and\
  \bibinfo {author} {\bibfnamefont {N.}~\bibnamefont {Trivedi}},\ }\href
  {\doibase 10.1103/PhysRevLett.104.066406} {\bibfield  {journal} {\bibinfo
  {journal} {Phys. Rev. Lett.}\ }\textbf {\bibinfo {volume} {104}},\ \bibinfo
  {pages} {066406} (\bibinfo {year} {2010})}\BibitemShut {NoStop}%
\bibitem [{\citenamefont {Sato}\ and\ \citenamefont
  {Tsunetsugu}(2016)}]{Sato2016}%
  \BibitemOpen
  \bibfield  {author} {\bibinfo {author} {\bibfnamefont {T.}~\bibnamefont
  {Sato}}\ and\ \bibinfo {author} {\bibfnamefont {H.}~\bibnamefont
  {Tsunetsugu}},\ }\href {\doibase 10.1103/PhysRevB.94.085110} {\bibfield
  {journal} {\bibinfo  {journal} {Phys. Rev. B}\ }\textbf {\bibinfo {volume}
  {94}},\ \bibinfo {pages} {085110} (\bibinfo {year} {2016})}\BibitemShut
  {NoStop}%
\bibitem [{\citenamefont {Ayral}\ and\ \citenamefont
  {Parcollet}(2015)}]{trilex1}%
  \BibitemOpen
  \bibfield  {author} {\bibinfo {author} {\bibfnamefont {T.}~\bibnamefont
  {Ayral}}\ and\ \bibinfo {author} {\bibfnamefont {O.}~\bibnamefont
  {Parcollet}},\ }\href {\doibase 10.1103/PhysRevB.92.115109} {\bibfield
  {journal} {\bibinfo  {journal} {Phys. Rev. B}\ }\textbf {\bibinfo {volume}
  {92}},\ \bibinfo {pages} {115109} (\bibinfo {year} {2015})}\BibitemShut
  {NoStop}%
\bibitem [{\citenamefont {Ayral}\ and\ \citenamefont
  {Parcollet}(2016{\natexlab{a}})}]{trilex2}%
  \BibitemOpen
  \bibfield  {author} {\bibinfo {author} {\bibfnamefont {T.}~\bibnamefont
  {Ayral}}\ and\ \bibinfo {author} {\bibfnamefont {O.}~\bibnamefont
  {Parcollet}},\ }\href {\doibase 10.1103/PhysRevB.93.235124} {\bibfield
  {journal} {\bibinfo  {journal} {Phys. Rev. B}\ }\textbf {\bibinfo {volume}
  {93}},\ \bibinfo {pages} {235124} (\bibinfo {year}
  {2016}{\natexlab{a}})}\BibitemShut {NoStop}%
\bibitem [{\citenamefont {Ayral}\ and\ \citenamefont
  {Parcollet}(2016{\natexlab{b}})}]{quadrilex}%
  \BibitemOpen
  \bibfield  {author} {\bibinfo {author} {\bibfnamefont {T.}~\bibnamefont
  {Ayral}}\ and\ \bibinfo {author} {\bibfnamefont {O.}~\bibnamefont
  {Parcollet}},\ }\href {\doibase 10.1103/PhysRevB.94.075159} {\bibfield
  {journal} {\bibinfo  {journal} {Phys. Rev. B}\ }\textbf {\bibinfo {volume}
  {94}},\ \bibinfo {pages} {075159} (\bibinfo {year}
  {2016}{\natexlab{b}})}\BibitemShut {NoStop}%
\bibitem [{\citenamefont {{Rohringer}}\ \emph {et~al.}(2017)\citenamefont
  {{Rohringer}}, \citenamefont {{Hafermann}}, \citenamefont {{Toschi}},
  \citenamefont {{Katanin}}, \citenamefont {{Antipov}}, \citenamefont
  {{Katsnelson}}, \citenamefont {{Lichtenstein}}, \citenamefont {{Rubtsov}},\
  and\ \citenamefont {{Held}}}]{Rohringer2017}%
  \BibitemOpen
  \bibfield  {author} {\bibinfo {author} {\bibfnamefont {G.}~\bibnamefont
  {{Rohringer}}}, \bibinfo {author} {\bibfnamefont {H.}~\bibnamefont
  {{Hafermann}}}, \bibinfo {author} {\bibfnamefont {A.}~\bibnamefont
  {{Toschi}}}, \bibinfo {author} {\bibfnamefont {A.~A.}\ \bibnamefont
  {{Katanin}}}, \bibinfo {author} {\bibfnamefont {A.~E.}\ \bibnamefont
  {{Antipov}}}, \bibinfo {author} {\bibfnamefont {M.~I.}\ \bibnamefont
  {{Katsnelson}}}, \bibinfo {author} {\bibfnamefont {A.~I.}\ \bibnamefont
  {{Lichtenstein}}}, \bibinfo {author} {\bibfnamefont {A.~N.}\ \bibnamefont
  {{Rubtsov}}}, \ and\ \bibinfo {author} {\bibfnamefont {K.}~\bibnamefont
  {{Held}}},\ }\href@noop {} {\bibfield  {journal} {\bibinfo  {journal} {ArXiv
  e-prints}\ } (\bibinfo {year} {2017})},\ \Eprint
  {http://arxiv.org/abs/1705.00024} {arXiv:1705.00024 [cond-mat.str-el]}
  \BibitemShut {NoStop}%
\bibitem [{Note1()}]{Note1}%
  \BibitemOpen
  \bibinfo {note} {See Supplemental Material for the staggered magnetization
  curves as a function of $U$ and $\delta $ for few temperatures; complementary
  data for the local DOS.}\BibitemShut {Stop}%
\bibitem [{\citenamefont {Georges}\ \emph {et~al.}(1996)\citenamefont
  {Georges}, \citenamefont {Kotliar}, \citenamefont {Krauth},\ and\
  \citenamefont {Rozenberg}}]{rmp}%
  \BibitemOpen
  \bibfield  {author} {\bibinfo {author} {\bibfnamefont {A.}~\bibnamefont
  {Georges}}, \bibinfo {author} {\bibfnamefont {G.}~\bibnamefont {Kotliar}},
  \bibinfo {author} {\bibfnamefont {W.}~\bibnamefont {Krauth}}, \ and\ \bibinfo
  {author} {\bibfnamefont {M.~J.}\ \bibnamefont {Rozenberg}},\ }\href {\doibase
  10.1103/RevModPhys.68.13} {\bibfield  {journal} {\bibinfo  {journal} {Rev.
  Mod. Phys.}\ }\textbf {\bibinfo {volume} {68}},\ \bibinfo {pages} {13}
  (\bibinfo {year} {1996})}\BibitemShut {NoStop}%
\bibitem [{\citenamefont {Fratino}\ \emph {et~al.}(2017)\citenamefont
  {Fratino}, \citenamefont {S\'emon}, \citenamefont {Charlebois}, \citenamefont
  {Sordi},\ and\ \citenamefont {Tremblay}}]{LorenzoAF2017}%
  \BibitemOpen
  \bibfield  {author} {\bibinfo {author} {\bibfnamefont {L.}~\bibnamefont
  {Fratino}}, \bibinfo {author} {\bibfnamefont {P.}~\bibnamefont {S\'emon}},
  \bibinfo {author} {\bibfnamefont {M.}~\bibnamefont {Charlebois}}, \bibinfo
  {author} {\bibfnamefont {G.}~\bibnamefont {Sordi}}, \ and\ \bibinfo {author}
  {\bibfnamefont {A.-M.~S.}\ \bibnamefont {Tremblay}},\ }\href {\doibase
  10.1103/PhysRevB.95.235109} {\bibfield  {journal} {\bibinfo  {journal} {Phys.
  Rev. B}\ }\textbf {\bibinfo {volume} {95}},\ \bibinfo {pages} {235109}
  (\bibinfo {year} {2017})}\BibitemShut {NoStop}%
\bibitem [{\citenamefont {Gull}\ \emph {et~al.}(2011)\citenamefont {Gull},
  \citenamefont {Millis}, \citenamefont {Lichtenstein}, \citenamefont
  {Rubtsov}, \citenamefont {Troyer},\ and\ \citenamefont {Werner}}]{millisRMP}%
  \BibitemOpen
  \bibfield  {author} {\bibinfo {author} {\bibfnamefont {E.}~\bibnamefont
  {Gull}}, \bibinfo {author} {\bibfnamefont {A.~J.}\ \bibnamefont {Millis}},
  \bibinfo {author} {\bibfnamefont {A.~I.}\ \bibnamefont {Lichtenstein}},
  \bibinfo {author} {\bibfnamefont {A.~N.}\ \bibnamefont {Rubtsov}}, \bibinfo
  {author} {\bibfnamefont {M.}~\bibnamefont {Troyer}}, \ and\ \bibinfo {author}
  {\bibfnamefont {P.}~\bibnamefont {Werner}},\ }\href {\doibase
  10.1103/RevModPhys.83.349} {\bibfield  {journal} {\bibinfo  {journal} {Rev.
  Mod. Phys.}\ }\textbf {\bibinfo {volume} {83}},\ \bibinfo {pages} {349}
  (\bibinfo {year} {2011})}\BibitemShut {NoStop}%
\bibitem [{\citenamefont {S\'emon}\ \emph
  {et~al.}(2014{\natexlab{a}})\citenamefont {S\'emon}, \citenamefont {Yee},
  \citenamefont {Haule},\ and\ \citenamefont {Tremblay}}]{patrickSkipList}%
  \BibitemOpen
  \bibfield  {author} {\bibinfo {author} {\bibfnamefont {P.}~\bibnamefont
  {S\'emon}}, \bibinfo {author} {\bibfnamefont {C.-H.}\ \bibnamefont {Yee}},
  \bibinfo {author} {\bibfnamefont {K.}~\bibnamefont {Haule}}, \ and\ \bibinfo
  {author} {\bibfnamefont {A.-M.~S.}\ \bibnamefont {Tremblay}},\ }\href
  {\doibase 10.1103/PhysRevB.90.075149} {\bibfield  {journal} {\bibinfo
  {journal} {Phys. Rev. B}\ }\textbf {\bibinfo {volume} {90}},\ \bibinfo
  {pages} {075149} (\bibinfo {year} {2014}{\natexlab{a}})}\BibitemShut
  {NoStop}%
\bibitem [{\citenamefont {S\'emon}\ and\ \citenamefont
  {Tremblay}(2012)}]{patrickCritical}%
  \BibitemOpen
  \bibfield  {author} {\bibinfo {author} {\bibfnamefont {P.}~\bibnamefont
  {S\'emon}}\ and\ \bibinfo {author} {\bibfnamefont {A.-M.~S.}\ \bibnamefont
  {Tremblay}},\ }\href {\doibase 10.1103/PhysRevB.85.201101} {\bibfield
  {journal} {\bibinfo  {journal} {Phys. Rev. B}\ }\textbf {\bibinfo {volume}
  {85}},\ \bibinfo {pages} {201101} (\bibinfo {year} {2012})}\BibitemShut
  {NoStop}%
\bibitem [{\citenamefont {S\'emon}\ \emph
  {et~al.}(2014{\natexlab{b}})\citenamefont {S\'emon}, \citenamefont {Sordi},\
  and\ \citenamefont {Tremblay}}]{patrickERG}%
  \BibitemOpen
  \bibfield  {author} {\bibinfo {author} {\bibfnamefont {P.}~\bibnamefont
  {S\'emon}}, \bibinfo {author} {\bibfnamefont {G.}~\bibnamefont {Sordi}}, \
  and\ \bibinfo {author} {\bibfnamefont {A.-M.~S.}\ \bibnamefont {Tremblay}},\
  }\href {\doibase 10.1103/PhysRevB.89.165113} {\bibfield  {journal} {\bibinfo
  {journal} {Phys. Rev. B}\ }\textbf {\bibinfo {volume} {89}},\ \bibinfo
  {pages} {165113} (\bibinfo {year} {2014}{\natexlab{b}})}\BibitemShut
  {NoStop}%
\bibitem [{\citenamefont {Mermin}\ and\ \citenamefont
  {Wagner}(1966)}]{MWtheorem}%
  \BibitemOpen
  \bibfield  {author} {\bibinfo {author} {\bibfnamefont {N.~D.}\ \bibnamefont
  {Mermin}}\ and\ \bibinfo {author} {\bibfnamefont {H.}~\bibnamefont
  {Wagner}},\ }\href {\doibase 10.1103/PhysRevLett.17.1133} {\bibfield
  {journal} {\bibinfo  {journal} {Phys. Rev. Lett.}\ }\textbf {\bibinfo
  {volume} {17}},\ \bibinfo {pages} {1133} (\bibinfo {year}
  {1966})}\BibitemShut {NoStop}%
\bibitem [{\citenamefont {Hohenberg}(1967)}]{Hohenberg:1967}%
  \BibitemOpen
  \bibfield  {author} {\bibinfo {author} {\bibfnamefont {P.~C.}\ \bibnamefont
  {Hohenberg}},\ }\href {\doibase 10.1103/PhysRev.158.383} {\bibfield
  {journal} {\bibinfo  {journal} {Phys. Rev.}\ }\textbf {\bibinfo {volume}
  {158}},\ \bibinfo {pages} {383} (\bibinfo {year} {1967})}\BibitemShut
  {NoStop}%
\bibitem [{\citenamefont {Schrieffer}\ \emph {et~al.}(1989)\citenamefont
  {Schrieffer}, \citenamefont {Wen},\ and\ \citenamefont {Zhang}}]{Zhang:1989}%
  \BibitemOpen
  \bibfield  {author} {\bibinfo {author} {\bibfnamefont {J.~R.}\ \bibnamefont
  {Schrieffer}}, \bibinfo {author} {\bibfnamefont {X.~G.}\ \bibnamefont {Wen}},
  \ and\ \bibinfo {author} {\bibfnamefont {S.~C.}\ \bibnamefont {Zhang}},\
  }\href {\doibase 10.1103/PhysRevB.39.11663} {\bibfield  {journal} {\bibinfo
  {journal} {Phys. Rev. B}\ }\textbf {\bibinfo {volume} {39}},\ \bibinfo
  {pages} {11663} (\bibinfo {year} {1989})}\BibitemShut {NoStop}%
\bibitem [{\citenamefont {Georges}\ and\ \citenamefont
  {Krauth}(1993)}]{GeorgesKrauthAFM:1993}%
  \BibitemOpen
  \bibfield  {author} {\bibinfo {author} {\bibfnamefont {A.}~\bibnamefont
  {Georges}}\ and\ \bibinfo {author} {\bibfnamefont {W.}~\bibnamefont
  {Krauth}},\ }\href {\doibase 10.1103/PhysRevB.48.7167} {\bibfield  {journal}
  {\bibinfo  {journal} {Phys. Rev. B}\ }\textbf {\bibinfo {volume} {48}},\
  \bibinfo {pages} {7167} (\bibinfo {year} {1993})}\BibitemShut {NoStop}%
\bibitem [{\citenamefont {Freericks}\ and\ \citenamefont
  {Jarrell}(1995)}]{FreericksJarrelAFM:1995}%
  \BibitemOpen
  \bibfield  {author} {\bibinfo {author} {\bibfnamefont {J.~K.}\ \bibnamefont
  {Freericks}}\ and\ \bibinfo {author} {\bibfnamefont {M.}~\bibnamefont
  {Jarrell}},\ }\href {\doibase 10.1103/PhysRevLett.74.186} {\bibfield
  {journal} {\bibinfo  {journal} {Phys. Rev. Lett.}\ }\textbf {\bibinfo
  {volume} {74}},\ \bibinfo {pages} {186} (\bibinfo {year} {1995})}\BibitemShut
  {NoStop}%
\bibitem [{\citenamefont {Hofstetter}\ and\ \citenamefont
  {Vollhardt}(1998)}]{hofstetter1998}%
  \BibitemOpen
  \bibfield  {author} {\bibinfo {author} {\bibfnamefont {W.}~\bibnamefont
  {Hofstetter}}\ and\ \bibinfo {author} {\bibfnamefont {D.}~\bibnamefont
  {Vollhardt}},\ }\href {\doibase 10.1002/andp.2060070105} {\bibfield
  {journal} {\bibinfo  {journal} {Annalen der Physik}\ }\textbf {\bibinfo
  {volume} {7}},\ \bibinfo {pages} {48} (\bibinfo {year} {1998})}\BibitemShut
  {NoStop}%
\bibitem [{\citenamefont {Chitra}\ and\ \citenamefont
  {Kotliar}(1999)}]{ChitraAFM:1999}%
  \BibitemOpen
  \bibfield  {author} {\bibinfo {author} {\bibfnamefont {R.}~\bibnamefont
  {Chitra}}\ and\ \bibinfo {author} {\bibfnamefont {G.}~\bibnamefont
  {Kotliar}},\ }\href {\doibase 10.1103/PhysRevLett.83.2386} {\bibfield
  {journal} {\bibinfo  {journal} {Phys. Rev. Lett.}\ }\textbf {\bibinfo
  {volume} {83}},\ \bibinfo {pages} {2386} (\bibinfo {year}
  {1999})}\BibitemShut {NoStop}%
\bibitem [{\citenamefont {Zitzler}\ \emph {et~al.}(2004)\citenamefont
  {Zitzler}, \citenamefont {Tong}, \citenamefont {Pruschke},\ and\
  \citenamefont {Bulla}}]{ZitzlerPruschkeAFM:2004}%
  \BibitemOpen
  \bibfield  {author} {\bibinfo {author} {\bibfnamefont {R.}~\bibnamefont
  {Zitzler}}, \bibinfo {author} {\bibfnamefont {N.-H.}\ \bibnamefont {Tong}},
  \bibinfo {author} {\bibfnamefont {T.}~\bibnamefont {Pruschke}}, \ and\
  \bibinfo {author} {\bibfnamefont {R.}~\bibnamefont {Bulla}},\ }\href
  {\doibase 10.1103/PhysRevLett.93.016406} {\bibfield  {journal} {\bibinfo
  {journal} {Phys. Rev. Lett.}\ }\textbf {\bibinfo {volume} {93}},\ \bibinfo
  {pages} {016406} (\bibinfo {year} {2004})}\BibitemShut {NoStop}%
\bibitem [{\citenamefont {Kanamori}(1963)}]{kanamori_electron_1963}%
  \BibitemOpen
  \bibfield  {author} {\bibinfo {author} {\bibfnamefont {J.}~\bibnamefont
  {Kanamori}},\ }\href {\doibase 10.1143/PTP.30.275} {\bibfield  {journal}
  {\bibinfo  {journal} {Progress of Theoretical Physics}\ }\textbf {\bibinfo
  {volume} {30}},\ \bibinfo {pages} {275} (\bibinfo {year} {1963})}\BibitemShut
  {NoStop}%
\bibitem [{\citenamefont {Brueckner}\ \emph {et~al.}(1960)\citenamefont
  {Brueckner}, \citenamefont {Soda}, \citenamefont {Anderson},\ and\
  \citenamefont {Morel}}]{Brueckner:1960}%
  \BibitemOpen
  \bibfield  {author} {\bibinfo {author} {\bibfnamefont {K.~A.}\ \bibnamefont
  {Brueckner}}, \bibinfo {author} {\bibfnamefont {T.}~\bibnamefont {Soda}},
  \bibinfo {author} {\bibfnamefont {P.~W.}\ \bibnamefont {Anderson}}, \ and\
  \bibinfo {author} {\bibfnamefont {P.}~\bibnamefont {Morel}},\ }\href
  {\doibase 10.1103/PhysRev.118.1442} {\bibfield  {journal} {\bibinfo
  {journal} {Phys. Rev.}\ }\textbf {\bibinfo {volume} {118}},\ \bibinfo {pages}
  {1442} (\bibinfo {year} {1960})}\BibitemShut {NoStop}%
\bibitem [{\citenamefont {Vilk}\ and\ \citenamefont
  {Tremblay}(1997)}]{Vilk:1997}%
  \BibitemOpen
  \bibfield  {author} {\bibinfo {author} {\bibfnamefont {Y.~M.}\ \bibnamefont
  {Vilk}}\ and\ \bibinfo {author} {\bibfnamefont {A.-M.~S.}\ \bibnamefont
  {Tremblay}},\ }\href@noop {} {\bibfield  {journal} {\bibinfo  {journal} {J.
  Phys I (France)}\ }\textbf {\bibinfo {volume} {7}},\ \bibinfo {pages} {1309 }
  (\bibinfo {year} {1997})}\BibitemShut {NoStop}%
\bibitem [{\citenamefont {Yang}\ \emph {et~al.}(2000)\citenamefont {Yang},
  \citenamefont {Lange},\ and\ \citenamefont {Kotliar}}]{kotliarSBaf}%
  \BibitemOpen
  \bibfield  {author} {\bibinfo {author} {\bibfnamefont {I.}~\bibnamefont
  {Yang}}, \bibinfo {author} {\bibfnamefont {E.}~\bibnamefont {Lange}}, \ and\
  \bibinfo {author} {\bibfnamefont {G.}~\bibnamefont {Kotliar}},\ }\href
  {\doibase 10.1103/PhysRevB.61.2521} {\bibfield  {journal} {\bibinfo
  {journal} {Phys. Rev. B}\ }\textbf {\bibinfo {volume} {61}},\ \bibinfo
  {pages} {2521} (\bibinfo {year} {2000})}\BibitemShut {NoStop}%
\bibitem [{\citenamefont {Camjayi}\ \emph {et~al.}(2006)\citenamefont
  {Camjayi}, \citenamefont {Chitra},\ and\ \citenamefont
  {Rozenberg}}]{albertoAF}%
  \BibitemOpen
  \bibfield  {author} {\bibinfo {author} {\bibfnamefont {A.}~\bibnamefont
  {Camjayi}}, \bibinfo {author} {\bibfnamefont {R.}~\bibnamefont {Chitra}}, \
  and\ \bibinfo {author} {\bibfnamefont {M.~J.}\ \bibnamefont {Rozenberg}},\
  }\href {\doibase 10.1103/PhysRevB.73.041103} {\bibfield  {journal} {\bibinfo
  {journal} {Phys. Rev. B}\ }\textbf {\bibinfo {volume} {73}},\ \bibinfo
  {pages} {041103} (\bibinfo {year} {2006})}\BibitemShut {NoStop}%
\bibitem [{\citenamefont {Bergeron}\ and\ \citenamefont
  {Tremblay}(2016)}]{DominicMEM}%
  \BibitemOpen
  \bibfield  {author} {\bibinfo {author} {\bibfnamefont {D.}~\bibnamefont
  {Bergeron}}\ and\ \bibinfo {author} {\bibfnamefont {A.-M.~S.}\ \bibnamefont
  {Tremblay}},\ }\href {\doibase 10.1103/PhysRevE.94.023303} {\bibfield
  {journal} {\bibinfo  {journal} {Phys. Rev. E}\ }\textbf {\bibinfo {volume}
  {94}},\ \bibinfo {pages} {023303} (\bibinfo {year} {2016})}\BibitemShut
  {NoStop}%
\bibitem [{\citenamefont {Moreo}\ \emph {et~al.}(1995)\citenamefont {Moreo},
  \citenamefont {Haas}, \citenamefont {Sandvik},\ and\ \citenamefont
  {Dagotto}}]{Moreo1995}%
  \BibitemOpen
  \bibfield  {author} {\bibinfo {author} {\bibfnamefont {A.}~\bibnamefont
  {Moreo}}, \bibinfo {author} {\bibfnamefont {S.}~\bibnamefont {Haas}},
  \bibinfo {author} {\bibfnamefont {A.~W.}\ \bibnamefont {Sandvik}}, \ and\
  \bibinfo {author} {\bibfnamefont {E.}~\bibnamefont {Dagotto}},\ }\href
  {\doibase 10.1103/PhysRevB.51.12045} {\bibfield  {journal} {\bibinfo
  {journal} {Phys. Rev. B}\ }\textbf {\bibinfo {volume} {51}},\ \bibinfo
  {pages} {12045} (\bibinfo {year} {1995})}\BibitemShut {NoStop}%
\bibitem [{\citenamefont {Preuss}\ \emph {et~al.}(1995)\citenamefont {Preuss},
  \citenamefont {Hanke},\ and\ \citenamefont {von~der Linden}}]{Preuss1995}%
  \BibitemOpen
  \bibfield  {author} {\bibinfo {author} {\bibfnamefont {R.}~\bibnamefont
  {Preuss}}, \bibinfo {author} {\bibfnamefont {W.}~\bibnamefont {Hanke}}, \
  and\ \bibinfo {author} {\bibfnamefont {W.}~\bibnamefont {von~der Linden}},\
  }\href {\doibase 10.1103/PhysRevLett.75.1344} {\bibfield  {journal} {\bibinfo
   {journal} {Phys. Rev. Lett.}\ }\textbf {\bibinfo {volume} {75}},\ \bibinfo
  {pages} {1344} (\bibinfo {year} {1995})}\BibitemShut {NoStop}%
\bibitem [{\citenamefont {Eskes}\ \emph {et~al.}(1991)\citenamefont {Eskes},
  \citenamefont {Meinders},\ and\ \citenamefont {Sawatzky}}]{eskes1991}%
  \BibitemOpen
  \bibfield  {author} {\bibinfo {author} {\bibfnamefont {H.}~\bibnamefont
  {Eskes}}, \bibinfo {author} {\bibfnamefont {M.~B.~J.}\ \bibnamefont
  {Meinders}}, \ and\ \bibinfo {author} {\bibfnamefont {G.~A.}\ \bibnamefont
  {Sawatzky}},\ }\href {\doibase 10.1103/PhysRevLett.67.1035} {\bibfield
  {journal} {\bibinfo  {journal} {Phys. Rev. Lett.}\ }\textbf {\bibinfo
  {volume} {67}},\ \bibinfo {pages} {1035} (\bibinfo {year}
  {1991})}\BibitemShut {NoStop}%
\bibitem [{\citenamefont {Wu}\ and\ \citenamefont {Lee}(2017)}]{wuAFevolution}%
  \BibitemOpen
  \bibfield  {author} {\bibinfo {author} {\bibfnamefont {H.-K.}\ \bibnamefont
  {Wu}}\ and\ \bibinfo {author} {\bibfnamefont {T.-K.}\ \bibnamefont {Lee}},\
  }\href {\doibase 10.1103/PhysRevB.95.035133} {\bibfield  {journal} {\bibinfo
  {journal} {Phys. Rev. B}\ }\textbf {\bibinfo {volume} {95}},\ \bibinfo
  {pages} {035133} (\bibinfo {year} {2017})}\BibitemShut {NoStop}%
\bibitem [{\citenamefont {Taranto}\ \emph {et~al.}(2012)\citenamefont
  {Taranto}, \citenamefont {Sangiovanni}, \citenamefont {Held}, \citenamefont
  {Capone}, \citenamefont {Georges},\ and\ \citenamefont
  {Toschi}}]{TarantoPRB2012}%
  \BibitemOpen
  \bibfield  {author} {\bibinfo {author} {\bibfnamefont {C.}~\bibnamefont
  {Taranto}}, \bibinfo {author} {\bibfnamefont {G.}~\bibnamefont
  {Sangiovanni}}, \bibinfo {author} {\bibfnamefont {K.}~\bibnamefont {Held}},
  \bibinfo {author} {\bibfnamefont {M.}~\bibnamefont {Capone}}, \bibinfo
  {author} {\bibfnamefont {A.}~\bibnamefont {Georges}}, \ and\ \bibinfo
  {author} {\bibfnamefont {A.}~\bibnamefont {Toschi}},\ }\href {\doibase
  10.1103/PhysRevB.85.085124} {\bibfield  {journal} {\bibinfo  {journal} {Phys.
  Rev. B}\ }\textbf {\bibinfo {volume} {85}},\ \bibinfo {pages} {085124}
  (\bibinfo {year} {2012})}\BibitemShut {NoStop}%
\bibitem [{\citenamefont {Mukuda}\ \emph {et~al.}(2008)\citenamefont {Mukuda},
  \citenamefont {Yamaguchi}, \citenamefont {Shimizu}, \citenamefont {Kitaoka},
  \citenamefont {Shirage},\ and\ \citenamefont {Iyo}}]{mukuda2008}%
  \BibitemOpen
  \bibfield  {author} {\bibinfo {author} {\bibfnamefont {H.}~\bibnamefont
  {Mukuda}}, \bibinfo {author} {\bibfnamefont {Y.}~\bibnamefont {Yamaguchi}},
  \bibinfo {author} {\bibfnamefont {S.}~\bibnamefont {Shimizu}}, \bibinfo
  {author} {\bibfnamefont {Y.}~\bibnamefont {Kitaoka}}, \bibinfo {author}
  {\bibfnamefont {P.}~\bibnamefont {Shirage}}, \ and\ \bibinfo {author}
  {\bibfnamefont {A.}~\bibnamefont {Iyo}},\ }\href {\doibase
  10.1143/JPSJ.77.124706} {\bibfield  {journal} {\bibinfo  {journal} {Journal
  of the Physical Society of Japan}\ }\textbf {\bibinfo {volume} {77}},\
  \bibinfo {pages} {124706} (\bibinfo {year} {2008})},\ \Eprint
  {http://arxiv.org/abs/http://dx.doi.org/10.1143/JPSJ.77.124706}
  {http://dx.doi.org/10.1143/JPSJ.77.124706} \BibitemShut {NoStop}%
\bibitem [{\citenamefont {Mukuda}\ \emph {et~al.}(2012)\citenamefont {Mukuda},
  \citenamefont {Shimizu}, \citenamefont {Iyo},\ and\ \citenamefont
  {Kitaoka}}]{mukuda2012}%
  \BibitemOpen
  \bibfield  {author} {\bibinfo {author} {\bibfnamefont {H.}~\bibnamefont
  {Mukuda}}, \bibinfo {author} {\bibfnamefont {S.}~\bibnamefont {Shimizu}},
  \bibinfo {author} {\bibfnamefont {A.}~\bibnamefont {Iyo}}, \ and\ \bibinfo
  {author} {\bibfnamefont {Y.}~\bibnamefont {Kitaoka}},\ }\href {\doibase
  10.1143/JPSJ.81.011008} {\bibfield  {journal} {\bibinfo  {journal} {Journal
  of the Physical Society of Japan}\ }\textbf {\bibinfo {volume} {81}},\
  \bibinfo {pages} {011008} (\bibinfo {year} {2012})}\BibitemShut {NoStop}%
\bibitem [{\citenamefont {Ino}\ \emph {et~al.}(1997)\citenamefont {Ino},
  \citenamefont {Mizokawa}, \citenamefont {Fujimori}, \citenamefont {Tamasaku},
  \citenamefont {Eisaki}, \citenamefont {Uchida}, \citenamefont {Kimura},
  \citenamefont {Sasagawa},\ and\ \citenamefont {Kishio}}]{InoMU}%
  \BibitemOpen
  \bibfield  {author} {\bibinfo {author} {\bibfnamefont {A.}~\bibnamefont
  {Ino}}, \bibinfo {author} {\bibfnamefont {T.}~\bibnamefont {Mizokawa}},
  \bibinfo {author} {\bibfnamefont {A.}~\bibnamefont {Fujimori}}, \bibinfo
  {author} {\bibfnamefont {K.}~\bibnamefont {Tamasaku}}, \bibinfo {author}
  {\bibfnamefont {H.}~\bibnamefont {Eisaki}}, \bibinfo {author} {\bibfnamefont
  {S.}~\bibnamefont {Uchida}}, \bibinfo {author} {\bibfnamefont
  {T.}~\bibnamefont {Kimura}}, \bibinfo {author} {\bibfnamefont
  {T.}~\bibnamefont {Sasagawa}}, \ and\ \bibinfo {author} {\bibfnamefont
  {K.}~\bibnamefont {Kishio}},\ }\href {\doibase 10.1103/PhysRevLett.79.2101}
  {\bibfield  {journal} {\bibinfo  {journal} {Phys. Rev. Lett.}\ }\textbf
  {\bibinfo {volume} {79}},\ \bibinfo {pages} {2101} (\bibinfo {year}
  {1997})}\BibitemShut {NoStop}%
\bibitem [{\citenamefont {Harima}\ \emph {et~al.}(2003)\citenamefont {Harima},
  \citenamefont {Fujimori}, \citenamefont {Sugaya},\ and\ \citenamefont
  {Terasaki}}]{HarimaMU}%
  \BibitemOpen
  \bibfield  {author} {\bibinfo {author} {\bibfnamefont {N.}~\bibnamefont
  {Harima}}, \bibinfo {author} {\bibfnamefont {A.}~\bibnamefont {Fujimori}},
  \bibinfo {author} {\bibfnamefont {T.}~\bibnamefont {Sugaya}}, \ and\ \bibinfo
  {author} {\bibfnamefont {I.}~\bibnamefont {Terasaki}},\ }\href {\doibase
  10.1103/PhysRevB.67.172501} {\bibfield  {journal} {\bibinfo  {journal} {Phys.
  Rev. B}\ }\textbf {\bibinfo {volume} {67}},\ \bibinfo {pages} {172501}
  (\bibinfo {year} {2003})}\BibitemShut {NoStop}%
\bibitem [{\citenamefont {Rietveld}\ \emph {et~al.}(1992)\citenamefont
  {Rietveld}, \citenamefont {Chen},\ and\ \citenamefont {van~der
  Marel}}]{RietveldMU}%
  \BibitemOpen
  \bibfield  {author} {\bibinfo {author} {\bibfnamefont {G.}~\bibnamefont
  {Rietveld}}, \bibinfo {author} {\bibfnamefont {N.~Y.}\ \bibnamefont {Chen}},
  \ and\ \bibinfo {author} {\bibfnamefont {D.}~\bibnamefont {van~der Marel}},\
  }\href {\doibase 10.1103/PhysRevLett.69.2578} {\bibfield  {journal} {\bibinfo
   {journal} {Phys. Rev. Lett.}\ }\textbf {\bibinfo {volume} {69}},\ \bibinfo
  {pages} {2578} (\bibinfo {year} {1992})}\BibitemShut {NoStop}%
\bibitem [{\citenamefont {van~der Marel}\ and\ \citenamefont
  {Rietveld}(1992)}]{dirkMU}%
  \BibitemOpen
  \bibfield  {author} {\bibinfo {author} {\bibfnamefont {D.}~\bibnamefont
  {van~der Marel}}\ and\ \bibinfo {author} {\bibfnamefont {G.}~\bibnamefont
  {Rietveld}},\ }\href {\doibase 10.1103/PhysRevLett.69.2575} {\bibfield
  {journal} {\bibinfo  {journal} {Phys. Rev. Lett.}\ }\textbf {\bibinfo
  {volume} {69}},\ \bibinfo {pages} {2575} (\bibinfo {year}
  {1992})}\BibitemShut {NoStop}%
\bibitem [{Note2()}]{Note2}%
  \BibitemOpen
  \bibinfo {note} {In these experiments, the DOS at $\omega =0$ vanishes while
  it in our case it is small. The inhomogeneity of the samples suggests that
  disorder can induce localisation effects at $\omega =0$, as pointed out in
  Ref.~\protect \rev@citealpnum {wuAFevolution}. The AF phase of cuprates at
  finite doping is generally considered metallic~\protect \cite
  {ando2001}.}\BibitemShut {Stop}%
\bibitem [{\citenamefont {Molegraaf}\ \emph {et~al.}(2002)\citenamefont
  {Molegraaf}, \citenamefont {Presura}, \citenamefont {van~der Marel},
  \citenamefont {Kes},\ and\ \citenamefont {Li}}]{Molegraaf2002}%
  \BibitemOpen
  \bibfield  {author} {\bibinfo {author} {\bibfnamefont {H.~J.~A.}\
  \bibnamefont {Molegraaf}}, \bibinfo {author} {\bibfnamefont {C.}~\bibnamefont
  {Presura}}, \bibinfo {author} {\bibfnamefont {D.}~\bibnamefont {van~der
  Marel}}, \bibinfo {author} {\bibfnamefont {P.~H.}\ \bibnamefont {Kes}}, \
  and\ \bibinfo {author} {\bibfnamefont {M.}~\bibnamefont {Li}},\ }\href
  {\doibase 10.1126/science.1069947} {\bibfield  {journal} {\bibinfo  {journal}
  {Science}\ }\textbf {\bibinfo {volume} {295}},\ \bibinfo {pages} {2239}
  (\bibinfo {year} {2002})}\BibitemShut {NoStop}%
\bibitem [{\citenamefont {Deutscher}\ \emph {et~al.}(2005)\citenamefont
  {Deutscher}, \citenamefont {Santander-Syro},\ and\ \citenamefont
  {Bontemps}}]{deutscher2005}%
  \BibitemOpen
  \bibfield  {author} {\bibinfo {author} {\bibfnamefont {G.}~\bibnamefont
  {Deutscher}}, \bibinfo {author} {\bibfnamefont {A.~F.}\ \bibnamefont
  {Santander-Syro}}, \ and\ \bibinfo {author} {\bibfnamefont {N.}~\bibnamefont
  {Bontemps}},\ }\href {\doibase 10.1103/PhysRevB.72.092504} {\bibfield
  {journal} {\bibinfo  {journal} {Phys. Rev. B}\ }\textbf {\bibinfo {volume}
  {72}},\ \bibinfo {pages} {092504} (\bibinfo {year} {2005})}\BibitemShut
  {NoStop}%
\bibitem [{\citenamefont {Carbone}\ \emph {et~al.}(2006)\citenamefont
  {Carbone}, \citenamefont {Kuzmenko}, \citenamefont {Molegraaf}, \citenamefont
  {van Heumen}, \citenamefont {Lukovac}, \citenamefont {Marsiglio},
  \citenamefont {van~der Marel}, \citenamefont {Haule}, \citenamefont
  {Kotliar}, \citenamefont {Berger}, \citenamefont {Courjault}, \citenamefont
  {Kes},\ and\ \citenamefont {Li}}]{PhysRevB.74.064510}%
  \BibitemOpen
  \bibfield  {author} {\bibinfo {author} {\bibfnamefont {F.}~\bibnamefont
  {Carbone}}, \bibinfo {author} {\bibfnamefont {A.~B.}\ \bibnamefont
  {Kuzmenko}}, \bibinfo {author} {\bibfnamefont {H.~J.~A.}\ \bibnamefont
  {Molegraaf}}, \bibinfo {author} {\bibfnamefont {E.}~\bibnamefont {van
  Heumen}}, \bibinfo {author} {\bibfnamefont {V.}~\bibnamefont {Lukovac}},
  \bibinfo {author} {\bibfnamefont {F.}~\bibnamefont {Marsiglio}}, \bibinfo
  {author} {\bibfnamefont {D.}~\bibnamefont {van~der Marel}}, \bibinfo {author}
  {\bibfnamefont {K.}~\bibnamefont {Haule}}, \bibinfo {author} {\bibfnamefont
  {G.}~\bibnamefont {Kotliar}}, \bibinfo {author} {\bibfnamefont
  {H.}~\bibnamefont {Berger}}, \bibinfo {author} {\bibfnamefont
  {S.}~\bibnamefont {Courjault}}, \bibinfo {author} {\bibfnamefont {P.~H.}\
  \bibnamefont {Kes}}, \ and\ \bibinfo {author} {\bibfnamefont
  {M.}~\bibnamefont {Li}},\ }\href {\doibase 10.1103/PhysRevB.74.064510}
  {\bibfield  {journal} {\bibinfo  {journal} {Phys. Rev. B}\ }\textbf {\bibinfo
  {volume} {74}},\ \bibinfo {pages} {064510} (\bibinfo {year}
  {2006})}\BibitemShut {NoStop}%
\bibitem [{\citenamefont {S\'{e}n\'{e}chal}\ and\ \citenamefont
  {Tremblay}(2004)}]{st}%
  \BibitemOpen
  \bibfield  {author} {\bibinfo {author} {\bibfnamefont {D.}~\bibnamefont
  {S\'{e}n\'{e}chal}}\ and\ \bibinfo {author} {\bibfnamefont {A.-M.~S.}\
  \bibnamefont {Tremblay}},\ }\href {\doibase 10.1103/PhysRevLett.92.126401}
  {\bibfield  {journal} {\bibinfo  {journal} {Phys. Rev. Lett.}\ }\textbf
  {\bibinfo {volume} {92}},\ \bibinfo {pages} {126401} (\bibinfo {year}
  {2004})}\BibitemShut {NoStop}%
\bibitem [{\citenamefont {Weber}\ \emph
  {et~al.}(2010{\natexlab{a}})\citenamefont {Weber}, \citenamefont {Haule},\
  and\ \citenamefont {Kotliar}}]{Weber:2010}%
  \BibitemOpen
  \bibfield  {author} {\bibinfo {author} {\bibfnamefont {C.}~\bibnamefont
  {Weber}}, \bibinfo {author} {\bibfnamefont {K.}~\bibnamefont {Haule}}, \ and\
  \bibinfo {author} {\bibfnamefont {G.}~\bibnamefont {Kotliar}},\ }\href
  {\doibase 10.1038/nphys1706} {\bibfield  {journal} {\bibinfo  {journal}
  {Nature Physics}\ }\textbf {\bibinfo {volume} {6}},\ \bibinfo {pages} {574}
  (\bibinfo {year} {2010}{\natexlab{a}})}\BibitemShut {NoStop}%
\bibitem [{\citenamefont {Weber}\ \emph
  {et~al.}(2010{\natexlab{b}})\citenamefont {Weber}, \citenamefont {Haule},\
  and\ \citenamefont {Kotliar}}]{cedricApical}%
  \BibitemOpen
  \bibfield  {author} {\bibinfo {author} {\bibfnamefont {C.}~\bibnamefont
  {Weber}}, \bibinfo {author} {\bibfnamefont {K.}~\bibnamefont {Haule}}, \ and\
  \bibinfo {author} {\bibfnamefont {G.}~\bibnamefont {Kotliar}},\ }\href
  {\doibase 10.1103/PhysRevB.82.125107} {\bibfield  {journal} {\bibinfo
  {journal} {Phys. Rev. B}\ }\textbf {\bibinfo {volume} {82}},\ \bibinfo
  {pages} {125107} (\bibinfo {year} {2010}{\natexlab{b}})}\BibitemShut
  {NoStop}%
\bibitem [{\citenamefont {Sordi}\ \emph {et~al.}(2010)\citenamefont {Sordi},
  \citenamefont {Haule},\ and\ \citenamefont {Tremblay}}]{sht}%
  \BibitemOpen
  \bibfield  {author} {\bibinfo {author} {\bibfnamefont {G.}~\bibnamefont
  {Sordi}}, \bibinfo {author} {\bibfnamefont {K.}~\bibnamefont {Haule}}, \ and\
  \bibinfo {author} {\bibfnamefont {A.-M.~S.}\ \bibnamefont {Tremblay}},\
  }\href {\doibase 10.1103/PhysRevLett.104.226402} {\bibfield  {journal}
  {\bibinfo  {journal} {Phys. Rev. Lett.}\ }\textbf {\bibinfo {volume} {104}},\
  \bibinfo {pages} {226402} (\bibinfo {year} {2010})}\BibitemShut {NoStop}%
\bibitem [{\citenamefont {Sordi}\ \emph {et~al.}(2011)\citenamefont {Sordi},
  \citenamefont {Haule},\ and\ \citenamefont {Tremblay}}]{sht2}%
  \BibitemOpen
  \bibfield  {author} {\bibinfo {author} {\bibfnamefont {G.}~\bibnamefont
  {Sordi}}, \bibinfo {author} {\bibfnamefont {K.}~\bibnamefont {Haule}}, \ and\
  \bibinfo {author} {\bibfnamefont {A.-M.~S.}\ \bibnamefont {Tremblay}},\
  }\href {\doibase 10.1103/PhysRevB.84.075161} {\bibfield  {journal} {\bibinfo
  {journal} {Phys. Rev. B}\ }\textbf {\bibinfo {volume} {84}},\ \bibinfo
  {pages} {075161} (\bibinfo {year} {2011})}\BibitemShut {NoStop}%
\bibitem [{\citenamefont {Sordi}\ \emph {et~al.}(2012)\citenamefont {Sordi},
  \citenamefont {S\'emon}, \citenamefont {Haule},\ and\ \citenamefont
  {Tremblay}}]{ssht}%
  \BibitemOpen
  \bibfield  {author} {\bibinfo {author} {\bibfnamefont {G.}~\bibnamefont
  {Sordi}}, \bibinfo {author} {\bibfnamefont {P.}~\bibnamefont {S\'emon}},
  \bibinfo {author} {\bibfnamefont {K.}~\bibnamefont {Haule}}, \ and\ \bibinfo
  {author} {\bibfnamefont {A.-M.~S.}\ \bibnamefont {Tremblay}},\ }\href
  {\doibase doi:10.1038/srep00547} {\bibfield  {journal} {\bibinfo  {journal}
  {Sci. Rep.}\ }\textbf {\bibinfo {volume} {2}},\ \bibinfo {pages} {547}
  (\bibinfo {year} {2012})}\BibitemShut {NoStop}%
\bibitem [{\citenamefont {Sordi}\ \emph {et~al.}(2013)\citenamefont {Sordi},
  \citenamefont {S\'emon}, \citenamefont {Haule},\ and\ \citenamefont
  {Tremblay}}]{sshtRHO}%
  \BibitemOpen
  \bibfield  {author} {\bibinfo {author} {\bibfnamefont {G.}~\bibnamefont
  {Sordi}}, \bibinfo {author} {\bibfnamefont {P.}~\bibnamefont {S\'emon}},
  \bibinfo {author} {\bibfnamefont {K.}~\bibnamefont {Haule}}, \ and\ \bibinfo
  {author} {\bibfnamefont {A.-M.~S.}\ \bibnamefont {Tremblay}},\ }\href
  {\doibase 10.1103/PhysRevB.87.041101} {\bibfield  {journal} {\bibinfo
  {journal} {Phys. Rev. B}\ }\textbf {\bibinfo {volume} {87}},\ \bibinfo
  {pages} {041101} (\bibinfo {year} {2013})}\BibitemShut {NoStop}%
\bibitem [{\citenamefont {Fratino}\ \emph
  {et~al.}(2016{\natexlab{a}})\citenamefont {Fratino}, \citenamefont {S\'emon},
  \citenamefont {Sordi},\ and\ \citenamefont {Tremblay}}]{LorenzoSC}%
  \BibitemOpen
  \bibfield  {author} {\bibinfo {author} {\bibfnamefont {L.}~\bibnamefont
  {Fratino}}, \bibinfo {author} {\bibfnamefont {P.}~\bibnamefont {S\'emon}},
  \bibinfo {author} {\bibfnamefont {G.}~\bibnamefont {Sordi}}, \ and\ \bibinfo
  {author} {\bibfnamefont {A.-M.~S.}\ \bibnamefont {Tremblay}},\ }\href
  {\doibase 10.1038/srep22715} {\bibfield  {journal} {\bibinfo  {journal} {Sci.
  Rep.}\ }\textbf {\bibinfo {volume} {6}},\ \bibinfo {pages} {22715} (\bibinfo
  {year} {2016}{\natexlab{a}})}\BibitemShut {NoStop}%
\bibitem [{\citenamefont {Fratino}\ \emph
  {et~al.}(2016{\natexlab{b}})\citenamefont {Fratino}, \citenamefont {S\'emon},
  \citenamefont {Sordi},\ and\ \citenamefont {Tremblay}}]{Lorenzo3band}%
  \BibitemOpen
  \bibfield  {author} {\bibinfo {author} {\bibfnamefont {L.}~\bibnamefont
  {Fratino}}, \bibinfo {author} {\bibfnamefont {P.}~\bibnamefont {S\'emon}},
  \bibinfo {author} {\bibfnamefont {G.}~\bibnamefont {Sordi}}, \ and\ \bibinfo
  {author} {\bibfnamefont {A.-M.~S.}\ \bibnamefont {Tremblay}},\ }\href
  {\doibase 10.1103/PhysRevB.93.245147} {\bibfield  {journal} {\bibinfo
  {journal} {Phys. Rev. B}\ }\textbf {\bibinfo {volume} {93}},\ \bibinfo
  {pages} {245147} (\bibinfo {year} {2016}{\natexlab{b}})}\BibitemShut
  {NoStop}%
\bibitem [{\citenamefont {Ando}\ \emph {et~al.}(2001)\citenamefont {Ando},
  \citenamefont {Lavrov}, \citenamefont {Komiya}, \citenamefont {Segawa},\ and\
  \citenamefont {Sun}}]{ando2001}%
  \BibitemOpen
  \bibfield  {author} {\bibinfo {author} {\bibfnamefont {Y.}~\bibnamefont
  {Ando}}, \bibinfo {author} {\bibfnamefont {A.~N.}\ \bibnamefont {Lavrov}},
  \bibinfo {author} {\bibfnamefont {S.}~\bibnamefont {Komiya}}, \bibinfo
  {author} {\bibfnamefont {K.}~\bibnamefont {Segawa}}, \ and\ \bibinfo {author}
  {\bibfnamefont {X.~F.}\ \bibnamefont {Sun}},\ }\href {\doibase
  10.1103/PhysRevLett.87.017001} {\bibfield  {journal} {\bibinfo  {journal}
  {Phys. Rev. Lett.}\ }\textbf {\bibinfo {volume} {87}},\ \bibinfo {pages}
  {017001} (\bibinfo {year} {2001})}\BibitemShut {NoStop}%
\end{thebibliography}

%


\onecolumngrid
\clearpage
\setcounter{figure}{0}

\begin{center}

{\bf Supplemental information: \\Effects of interaction strength, doping, and frustration on the antiferromagnetic phase of the two-dimensional Hubbard model}

\vspace{5mm}
{L. Fratino, M. Charlebois, P. S\'emon, G. Sordi, and A.-M. S. Tremblay}

\end{center}

\vspace{5mm}

\begin{figure}[h!]
\centering{
\includegraphics[width=0.65\linewidth]{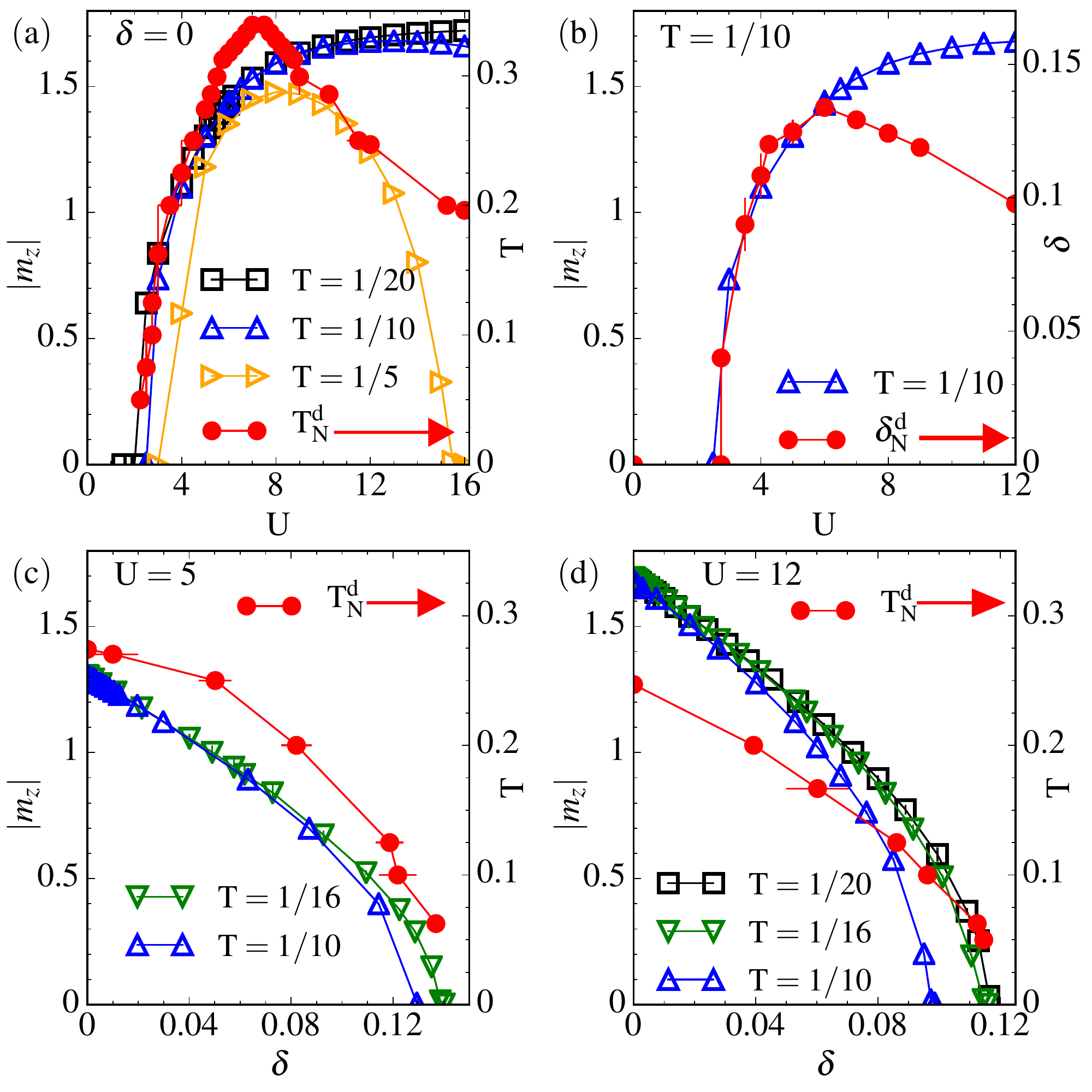}
}
\caption{(a) Data in all panels are shown for $t'=-0.1$. Staggered magnetization $m_z= \frac{2}{N_c} \sum_i(-1)^i (n_{i\uparrow} - n_{i\downarrow})$ as a function of $U$ for $\delta=0$. Data are shown for $T=1/20$ (black squares), $T=1/10$ (blue up triangles) and $T=1/5$ (orange right triangles). At the lowest temperature, $m_z(U)$ does not correlate with $T_N^d(U)$ (full red circles). 
(b) $m_z$ as a function of $U$ for $\delta=0$ and $T=1/10$. Note that $m_z(U)|_{T=1/10}$ does not correlate with $\delta_N^d(U)|_{T=1/10}$ (full red circles). 
(c) $m_z$ as a function of $\delta$ for $U=5$, and temperatures $T=1/16$ (green down triangles) and $T=1/10$ (blue up triangles). $m_z(\delta)$ correlates with $T_N^d(\delta)$ (full red circles). 
(d) $m_z$ as a function of $\delta$ for $U=12$, and temperatures $T=1/20$ (black squares), $T=1/16$ (green down triangles) and $T=1/10$ (blue up triangles). $m_z(\delta)$ correlates with $T_N^d(\delta)$ (full red circles).
The antiferromagnetic to paramagnetic phase boundary is obtained from the mean of the two values of temperature, doping or $U$ where $m_z$ changes from finite to a small value (here $m_z$= 0.045). This is consistent with a second-order transition at $\delta_c$.
}
\label{figSM1}
\end{figure}

\begin{figure}[h!]
\centering{
\includegraphics[width=0.60\linewidth]{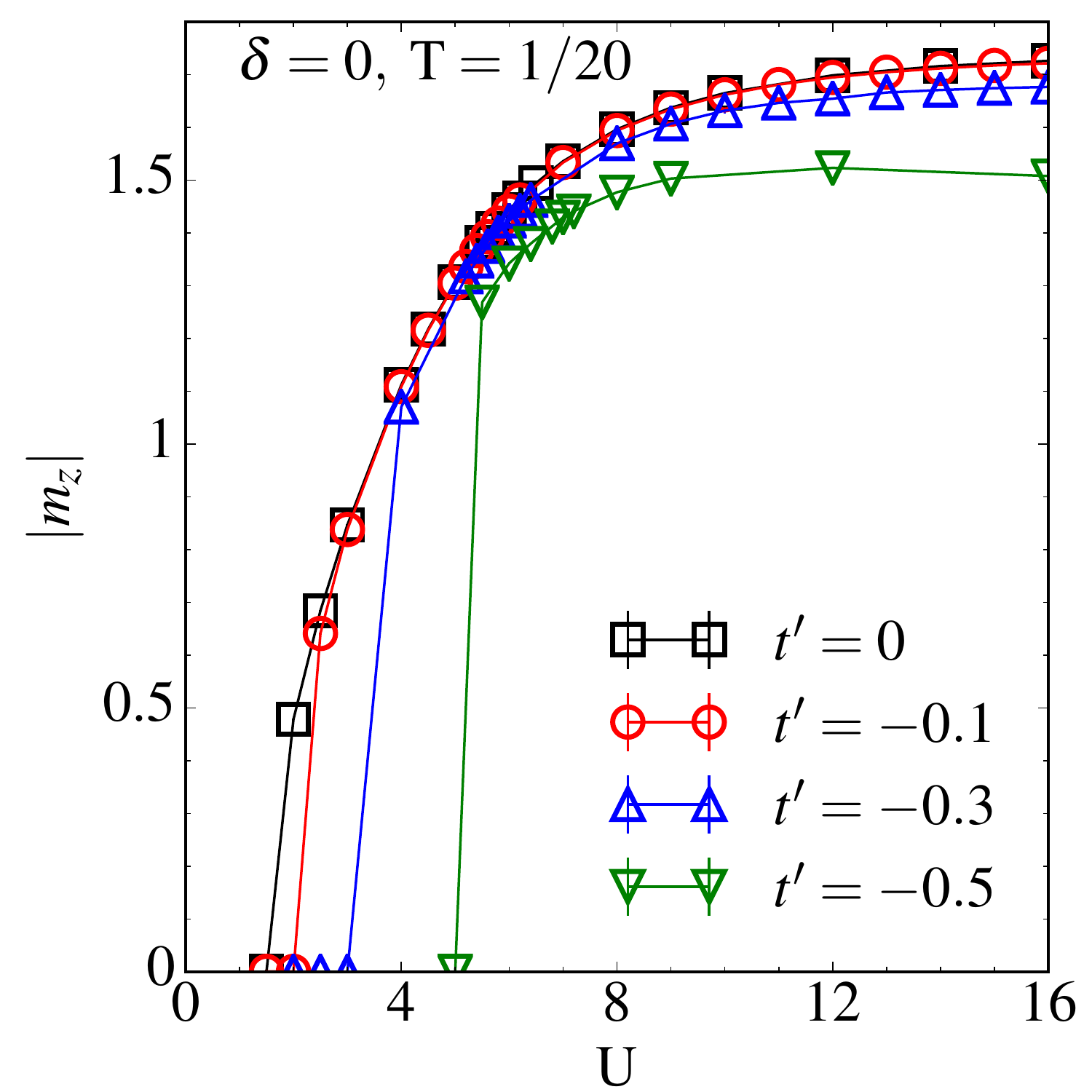}
}
\caption{Staggered magnetization $m_z$ as a function of $U$ for $\delta=0$ and $T=1/20$.  Data are shown for $t'=0$ (black squares), $t'=-0.1$ (red circles), $t'=-0.3$ (blue up triangles) and $t'=-0.5$ (green down triangles). There is a jump in $m_z$ as a function of $U$ for finite $t'$, consistent with a first-order transition at $U_c$. }
\label{figSM2}
\end{figure}

\begin{figure}[h!]
\centering{
\includegraphics[width=0.80\linewidth]{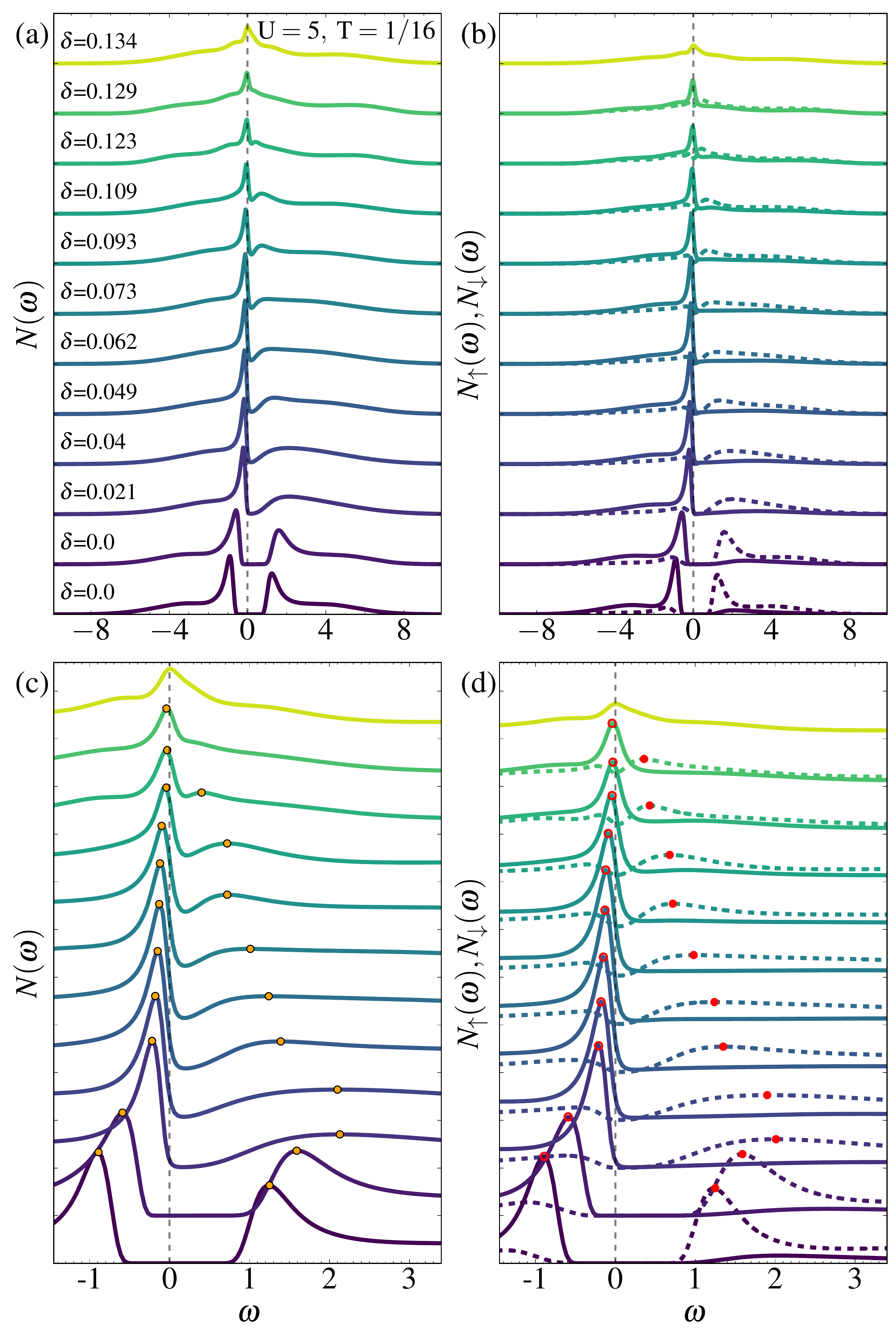}
}
\caption{(a) $N(\omega)$ in the antiferromagnetic state and (b) its two spin projections $N_\uparrow(\omega)$, $N_\downarrow(\omega)$, for $U=5$, $t'=-0.1$, $T=1/16$ and different values of doping $\delta$. The vertical dashed line at $\omega=0$ indicates the Fermi level. (c), (d) Zoom on the low frequency part of the data in panels (a) and (b), respectively. In panel (c), the full orange circles indicate the maximum of the Bogoliubov peaks. The polarisation of these peaks is sizeable, as can be seen in panel (d) from the maxima in $N_\uparrow (\omega)$, $N_\downarrow (\omega)$ (full and open circles, respectively). This figure extends Figures 3(a), 3(b) of the main text, where fewer values of doping are shown. 
}
\label{figSM3}
\end{figure}

In this supplemental information, we first present in Fig.~\ref{figSM1} the raw data of the staggered magnetization $m_z$, to complement the colormaps in Figure~1 of the main text. We also show in Fig.~\ref{figSM2} the raw data of $m_z$ to complement the antiferromagnetic boundaries of the Figure 2 of the main text. Finally, Fig.~\ref{figSM3} and Fig.~\ref{fig4} display the density of states for different values of hole doping $\delta$, to extend data shown in Figures~3(a), 3(b), 3(e), 3(f) of the main text.

\vspace{1cm}

\begin{figure}[h!]
\centering{
\includegraphics[width=0.80\linewidth]{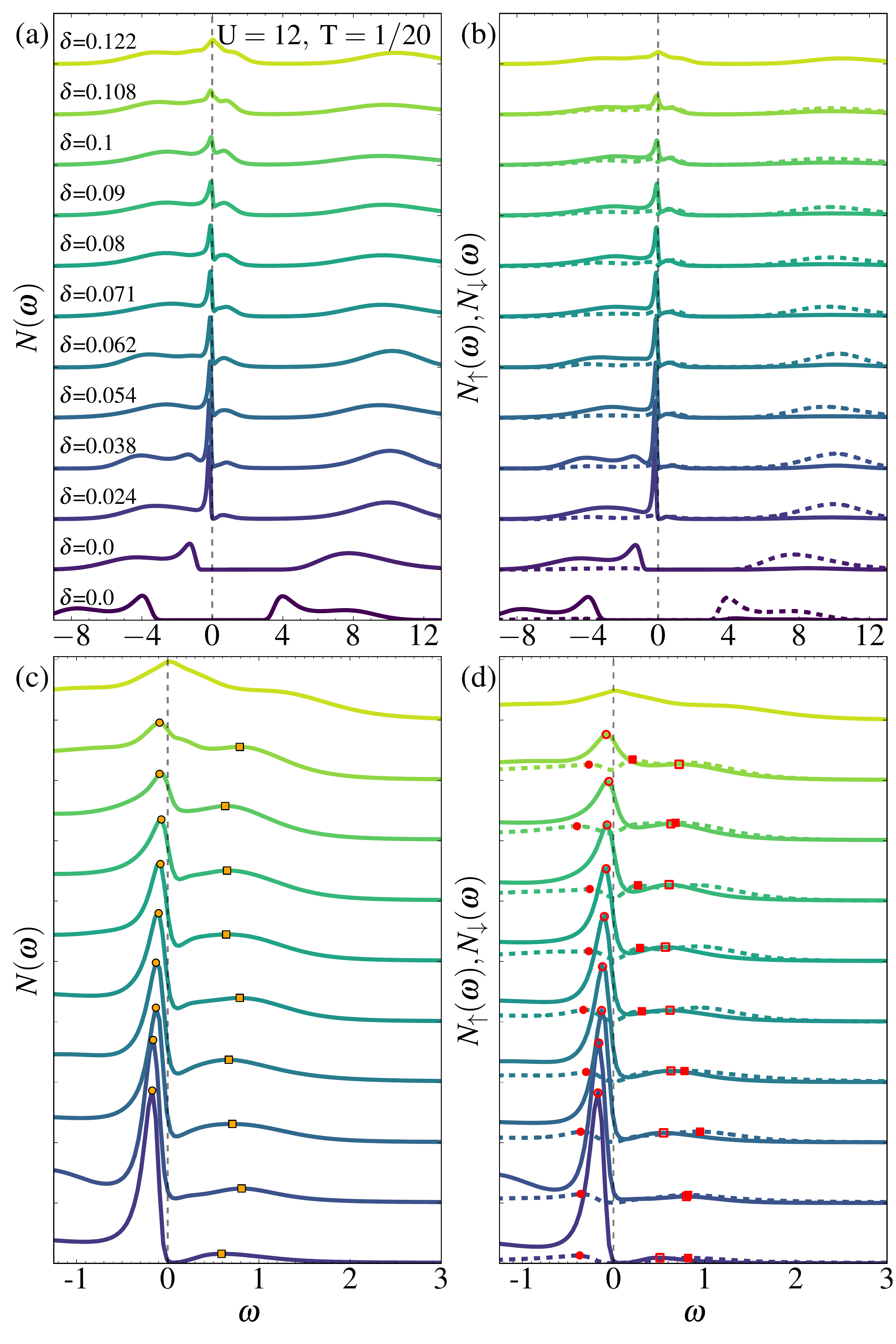}
}
\caption{a) $N(\omega)$ in the antiferromagnetic state and (b) its two spin projections $N_\uparrow(\omega)$, $N_\downarrow(\omega)$, for $U=12$, $t'=-0.1$, $T=1/20$ and different values of doping $\delta$. The vertical dashed line at $\omega=0$ indicates the Fermi level. (c), (d) Zoom on the low frequency part of the data in panels (a) and (b), respectively. Full orange circles indicate the maximum of the lower Bogoliubov peak. Orange squares indicate the maximum of the ``in-gap'' states. The lower Bogoliubov peak is almost fully spin polarized, as can be seen in panel (d) from the maxima in $N_\uparrow (\omega)$, $N_\downarrow (\omega)$ (full and open circles, respectively). In sharp contrast, the ``in-gap'' states are only slightly spin polarized, as suggested by the comparable maxima in $N_\uparrow (\omega)$, $N_\downarrow (\omega)$ (full and open squares, respectively). This figure extends Figures 3(e), 3(f) of the main text, where fewer values of doping are shown. 
}
\label{figSM4}
\end{figure}

\end{document}